\documentclass[aps,pra,twocolumn,showpacs,letterpaper,tighten,float,superscriptaddress,nofootinbib,reprint]{revtex4-1}
\usepackage{amsmath}
\usepackage[mathscr]{euscript}
\usepackage{gensymb,amsfonts}
\usepackage{amssymb}
\usepackage{array}
\usepackage[T1]{fontenc}
\usepackage{lmodern}
\usepackage{tabularx}
\usepackage{fancyhdr}
\usepackage{graphicx}
\graphicspath{{images/}}
\usepackage[english]{babel}
\usepackage{amsthm,mathtools}
\usepackage{braket}
\usepackage{tikz}
\usepackage{amsbsy}
\usepackage{mathrsfs}
\usepackage{epsfig}
\usepackage{textcomp}
\usepackage{color}
\usepackage{bm,hyperref}
\usepackage[titletoc,toc,title]{appendix}
\usepackage[normalem]{ulem}
\usepackage[splitrule]{footmisc}

\newcommand{\R}{\mathbb{R}}

\newcommand{\Hi}{\mathcal{H}}
\newcommand{\outerp}[2]{\ket{#1}\!\bra{#2}}
\newcommand{\hlf}{\frac{1}{2}}

\newcommand{\prtl}[2]{\frac{\partial #1}{\partial #2}}

\newcommand{\mc}{\mathcal}
\newcommand{\mb}{\mathbb}
\newcommand{\supp}{\text{Supp}}

\newcommand{\bk}{\mathbf{k}}
\newcommand{\mbf}{\mathbf}
\newcommand{\ad}{\text{Ad}}

\newcommand{\beq}{\begin{equation}}
\newcommand{\eeq}{\end{equation}}

\begin{document}

\title{Entanglement Spectra of Stabilizer Codes:\\
A Window into Gapped Quantum Phases of Matter
}

\author{Albert T. Schmitz} 
\affiliation{
Department of Physics and Center for Theory of Quantum Matter,
University of Colorado, Boulder, Colorado 80309, USA
}

\author{Sheng-Jie Huang}
\affiliation{
Department of Physics and Center for Theory of Quantum Matter,
University of Colorado, Boulder, Colorado 80309, USA
}

\author{Abhinav Prem}
\affiliation{
Princeton Center for Theoretical Science,
Princeton University, NJ 08544, USA
}

\begin{abstract}
The entanglement spectrum (ES) provides a barometer of quantum entanglement and encodes physical information beyond that contained in the entanglement entropy. In this paper, we explore the ES of stabilizer codes, which furnish exactly solvable models for a plethora of gapped quantum phases of matter. Studying the ES for stabilizer Hamiltonians in the presence of arbitrary weak local perturbations thus allows us to develop a general framework within which the entanglement features of gapped topological phases can be computed and contrasted. In particular, we study models harboring fracton order, both type-I and type-II, and compare the resulting ES with that of both conventional topological order and of (strong) subsystem symmetry protected topological (SSPT) states. We find that \textit{non-local surface stabilizers} (NLSS), a set of symmetries of the Hamiltonian which form on the boundary of the entanglement cut, act as purveyors of universal non-local features appearing in the entanglement spectrum. While in conventional topological orders and fracton orders, the NLSS retain a form of topological invariance with respect to the entanglement cut, subsystem symmetric systems---fracton and SSPT phases---additionally show a non-trivial geometric dependence on the entanglement cut, corresponding to the subsystem symmetry. This sheds further light on the interplay between geometric and topological effects in fracton phases of matter and demonstrates that strong SSPT phases harbour a measure of quasi-local entanglement beyond that encountered in conventional SPT phases. We further show that a version of the edge-entanglement correspondence, established earlier for gapped two-dimensional topological phases, also holds for gapped three-dimensional fracton models. 
\end{abstract}

\maketitle

%%%%%%%%%%%%%%%%%%%%%%%%%%%%%%%%%%%%%%%%
\tableofcontents

\section{Introduction}
\label{intro}

The study of interacting many-body quantum phases of matter has been greatly influenced and aided by developments in the field of quantum information; indeed, it is now widely appreciated that quantum entanglement plays a vital role in the characterisation of a wide variety of zero temperature complex many-body systems. Entanglement has proven a particularly efficacious tool in the study of \textit{gapped} quantum phases, especially in lower spatial dimensions ($d<3$) where it has provided a classification of gapped 1d phases~\cite{chen2010,fidkowski2011} and a powerful diagnostic for topological order in $d=2$ spatial dimensions~\cite{preskill,levin2006}. Gapped topologically ordered phases of matter are believed to be entirely characterised by the universal properties of their ground states, and a defining feature of these phases, which evade description in terms of any local order parameter, is the pattern of long range entanglement (LRE) in their ground state(s) (see e.g. Ref.~\cite{groverreview} for a review). A potent and oft utilised probe for the presence of LRE is the entanglement entropy---for topologically ordered systems in two spatial dimensions, the entanglement entropy contains a universal subleading `constant' term, known as the ``topological entanglement entropy (TEE),'' which partially characterises such phases and is intimately linked to their topological quantum field theory (TQFT) description. 

The \textit{entanglement spectrum} (ES), introduced by Li and Haldane in the context of fractional quantum Hall fluids~\cite{lihaldane}, provides a more general probe of quantum entanglement and encodes physical information beyond that contained in the entanglement entropy. For a given bipartition of a system into two regions $A$ and $B$, the ``entanglement Hamiltonian'' $H_A$ of its ground state $\ket{\Psi}$ is defined through $\rho_A \equiv e^{-H_A}/Z$, where $Z = \text{Tr}\,e^{-H_A}$ and $\rho_A = \text{Tr}_B \outerp{\Psi}{\Psi}$ is the reduced density matrix defined on region $A$. Since the entanglement cut mimics a physical boundary for the system, $H_A$ can be crudely understood as the Hamiltonian for a physical edge of the system; based on this observation, Li and Haldane conjectured that the ES contains universal information about the low-energy boundary excitations \textit{i.e.,} the ground state wave function in the bulk encodes information about dynamics at the edge. 

Following the original proposal, the entanglement spectrum has been widely used to identify and distinguish quantum phases of matter, and has been especially fruitful in characterising gapped topological phases. While much of the initial work focused on chiral topological orders in 2d, including fractional quantum Hall fluids~\cite{regnault1,papic,chandran2011,dubail2012,cano2015,qi2012}, this was later extended to include symmetry protected topological (SPT) states~\cite{swingle2012,prodan2010,pollmann2010,turner2010,fidkowski2010,alba2012,choo2018} as well as topologically ordered states with gapped boundaries~\cite{ho1,ho2,berg2017,stringnetespec}. In all cases, a key result is the existence of a correspondence between the low-lying spectrum of the physical edge states and the low-lying entanglement spectrum, often referred to as an edge-ES correspondence.

In three spatial dimensions, fracton order (see Ref.~\cite{fractonreview} for a review) has emerged as a new platform for realising long-range entanglement in gapped quantum phases\footnote{\textit{Gapless} fracton order remains an equally active area of research, where symmetric tensor gauge theories have emerged as a powerful formalism within which to encapsulate much of the fracton phenomena~\cite{sub,genem,prem2,gromov,han3,bulmash}---we will not discuss these here.}, exhibiting features both familiar and distinct from those encountered in conventional topological order. Familiar features include the presence of long-range entanglement, locally indistinguishable ground states on non-trivial manifolds, and topologically charged excitations which cannot be created locally. The crucial distinguishing feature of these phases---discovered first in a series of exactly solvable models~\cite{chamon,haah,yoshida,fracton1,fracton2}---which has engendered much recent  activity~\cite{williamson,slagle1,prem,hsieh,leomichael,pai,shirleyfrac,twisted,pai2,supersolid,potter,yizhilego}, is that the mobility of certain excitations (fractons) is strictly verboten, while that of other excitations (sub-dimensional particles) is restricted along sub-dimensional manifolds of the three dimensional lattice. Unlike topologically ordered phases, whose ground state degeneracy depends only on the topology of the underlying manifold, for fracton phases the number of ground states grows sub-extensively, revealing their sensitivity to the underlying geometry. In this precise sense, fracton phases differ from topologically ordered phases in that they do not admit a low-energy description as a TQFT; indeed, a series of recent works have emphasised the \textit{geometric} nature of fracton order~\cite{slagle2,slagle3,shirleygeneral,cagenet,symmetric,yanholography,foliatedfield,gromov2}. 

Concurrently, the field of subsystem symmetry protected topological (SSPT) phases~\cite{yizhi1,devakulfractal,strongSSPT,kubica,stephen2018,spurious} has emerged as a close relative of fracton order, with a large number of fracton models obtained through gauging the subsystem symmetry of various SSPT phases~\cite{fracton2,yizhi1,yizhi2,shirleygauging,williamsonSET}. Subsystem symmetries, alternatively referred to as ``gauge-like'' symmetries~\cite{ortiz}, act on rigid subsystems which cannot be deformed, such as rigid planes of the cubic lattice. Besides their deep connection to fracton phases, it has also been realised that $d=2$ SSPT phases protected by rigid line-like symmetries can act as a universal resource for measurement-based quantum computation~\cite{elsecluster,devakulMBQC,stephen2018}. 

Given the theoretical discovery of these novel gapped phases, it is natural to ask whether measures of entanglement, such as the entanglement entropy or entanglement spectrum, exhibit features distinct from those seen in SPT phases or those with topological order. Early work in this direction has mostly focused on understanding the EE of these phases, with their ES remaining largely uncharted~\cite{decipher,regnault2,han2,albert,chenfoliatedent,spurious,yanholography}. In this paper, we aim to fill this lacuna by studying the entanglement spectra of stabilizer codes, which encode a large class of gapped quantum phases of matter. As such, stabilizer codes provide a complementary language to that of TQFTs, furnishing zero correlation length Hamiltonians which are sums of commuting projectors and which describe an exactly solvable point within some phase. Importantly, stabilizer codes provide a universal language within which we can compare and contrast the entanglement structure of fracton and SSPT phases with that of topologically ordered states. 

Specifically, in this paper, we develop a general procedure for deriving the ES for the ground state of a stabilizer code Hamiltonian in the presence of arbitrary weak, local perturbations. Typically, when calculating the TEE in the ground state of some stabilizer Hamiltonian, one first derives the EE for the fixed point Hamiltonian and then argues that any contributions to the TEE are invariant under local perturbations. However, perturbations play a crucial role in revealing the structure of the ES and must be taken into account from the beginning. Indeed, while the ES is flat for any unperturbed stabilizer code, it presents universal features in the presence of weak, local perturbations for all models considered here.

As examples of topological order, we consider the $d=2,3$ toric code models and find that their entanglement Hamiltonians (EH) universally map onto $\mb{Z}_2$ invariant $d=1,2$ Ising models acting on effective spin-$1/2$ variables. As a consequence of this, we recover a version of the edge-ES correspondence first established in Ref.~\cite{ho1}. We then consider the X-cube model~\cite{fracton2} and Haah's cubic code~\cite{haah} as examples of type-I and type-II fracton orders respectively and find that the geometric nature of these phases is manifest in their ES as well. That is, the EH for these models can be mapped onto an effective subsystem symmetric Ising-like model only for those entanglement cuts consistent with the planar (fractal) subsystem symmetries of the X-cube (cubic code). Thus, we show that the ES serves as a clear entanglement measure distinguishing fracton order from topological order, adding to the existing diagnostics for fracton order~\cite{devakul,pinch,albert}. We also provide strong evidence for a correspondence between the low-lying ES and the low-lying spectrum of physical edge states, extending the validity of the edge-ES correspondence to gapped fracton phases. Finally, we consider the $d=2$ cluster state as our example of a strong SSPT phase and argue that the so-called ``spurious'' contributions to the TEE, found in Ref.~\cite{spurious}, are in fact evidence of \textit{quasi-local entanglement} in such phases. In other words, we find that the entanglement structure of this system harbours features redolent of LRE systems which, despite not being topological in nature, remain robust against all subsystem symmetric perturbations. Thus, we establish SSPT phases as lying between conventional SRE SPT phases and LRE topological orders, thereby expanding the dictionary of possible patterns of entanglement in gapped phases.

The balance of this paper is organised as follows: in Section~\ref{ES} we start by reviewing some basic notions of the stabilizer formalism pertinent to this paper. In Section~\ref{sec:unpert}, we derive the flat entanglement spectra of any stabilizers' eigenstate and discuss some consequences, including a review of {\it recoverable information} and {\it emergent Gauss' laws},  as well as of the Schmidt decomposition for the stabilizers' eigenstates. This is followed by the perturbative analysis in Section~\ref{sec:pert}, where we arrive at the central result of the paper: a general formula for the entanglement Hamiltonian of a stabilizer ground state in the presence of an arbitrary weak, local perturbation for any bipartition of the qubits $(A, B)$, with $B=A^c$. In Section~\ref{sec:edge-es}, we apply our general formalism to gapped LRE phases: the $d=2,3$ toric codes, the X-cube model, and Haah's cubic code, with the latter two being examples of type-I and type-II fracton models, respectively. Section~\ref{sec:sspt} applies our general method to a strong-SSPT model, namely the $d=2$ cluster state, where we discuss signatures of subsystem symmetries in the entanglement and introduce the notion of quasi-local entanglement. Section~\ref{sec:summary} summarises the relation between all models considered herein and discusses a conjecture which extends our results to low-energy excited states. Lastly, we make our final remarks in Section~\ref{cncls}.   

%%%%%%%%%%%%%%%%%%%%%%%%%%%%%%%%%%%%%%%%

\section{Entanglement Spectra of Stabilizer Code Hamiltonians}
\label{ES}

We begin this section by reviewing some basics regarding the stabilizer formalism before deriving general results for the structure of quantum entanglement in these systems. In certain cases, we rely upon results previously established in Refs.~\cite{Haah2013,han2,albert}, and provide details only when required for the remainder. After reviewing the requisite background, we present a general method for deriving the entanglement spectrum of the ground state of a stabilizer Hamiltonian in the presence of arbitrary perturbations, allowing us to compare and contrast a myriad of phases within the same framework. From the ES, one can also extract the entanglement entropy and recoverable information~\cite{albert}; however, as the EE has been studied for both fracton order and SSPTs before, we only comment on this briefly. 

\subsection{The Stabilizer Formalism}

The stabilizer formalism provides a common language within which a large class of quantum many-body systems and quantum error-correcting codes can be efficiently described (see Ref.~\cite{finiteTrmp} for a review). It provides a unifying framework for studying several interacting quantum many-body systems, as certain exactly solvable points within some gapped phase can be described in terms of a Hamiltonian which is the sum of commuting Pauli operators---stabilizers. Each of the models considered here will admit such a description.

The models under consideration here consist of a set of $N$ local qubits (spin-1/2 degrees of freedom) living on the edges or vertices of a simple graph, with $q$ qubits per each edge or vertex, such that $N = q |V|$, where $V$ is the graph set. Here, $|\cdots|$ denotes the size of a finite set. The Hilbert space of the system $\Hi \simeq \mb{C}_2^{\otimes N}$ is given by the product space of all qubits. 

The \textit{Pauli group} $P$ acting on $N$ qubits is defined as the set of all Pauli operators---with individual qubit Pauli operators denoted by $\{X,Y,Z\}$---acting on the qubits, modulo any phase of $\pm 1,\pm i$. The \textit{stabilizer set} $\mc{S} \subset P$ is a subset of the Pauli group comprised of mutually commuting operators, which satisfies $|\mc S| \geq N$ (there are at least as many stabilizers as qubits), $-I_{\Hi} \not\in \mc{S}$ (all of $\mc{S}$ can have a positive eigenvalue), and $\supp(\mc{S}) = V$ (each element of $V$ is acted upon non-trivially by at least one stabilizer in $\mc{S}$). The set of states $\ket{\psi}$ which are stabilized by $\mc{S} =\{O_s\}$:
\beq
O_s \ket{\psi} = \ket{\psi},
\eeq
for all $O_s \in \mc{S}$, form the ground state manifold for the stabilizer code Hamiltonian
\beq
\label{eq:stab}
H_{\mc{S}} = -\sum_s J_s O_s, \quad (J_s > 0)
\eeq
\textit{i.e.,} states invariant under the action of elements of $\mc{S}$ span the ground state subspace of the Hamiltonian $H_{\mc{S}}$.

Since all members of $\mc{S}$ commute, stabilizers in $\mc{S}$ multiplicatively generate an Abelian group $G = \{\prod_{s \in F} O_s: F \in \mathbb{P}[\mc{S}] \}$, where the power set of $\mc{S}$, denoted $\mathbb{P}[\mc{S}]$, is the set of all subsets of $\mc{S}$. Not all stabilizers are independent, so $\mc S$ may over-determine $G$. We refer to any $C \subseteq \mc S$ such that $\prod_{s\in C} O_s = I$ as a {\it constraint}. Following \cite{Haah2013, han2, albert}, let $d_G =\log_2|G|$ and $\{O_i\}_{i \leq d_G} \subseteq \mc{S}$ be a complete, independent generating set for $G$. Elements $g \in G$ can hence be labelled by a binary vector $\mathbf{n} = (n_1,n_2,\dots,n_{d_G}) \in \{0,1\}^{d_G}$ %, with $n_i \in \{0,1\}$,
 through
\beq
g(\mathbf{n}) = \prod_{i=1}^{d_G} O_i^{n_i}.
\eeq
With every vector $\bk \in \{0,1\}^{d_{G}}$, we associate a projection operator
\begin{align}
\label{eq:proj}
\mc{P}^{\bk} = \frac{1}{|G|} \sum_{g(\mathbf{n}) \in G} (-1)^{\bk\cdot \mathbf{n}} g(\mathbf{n}),
\end{align}
where $\bk \cdot \mathbf{n}$ is the binary dot product of $\bk$ and $\mathbf{n}$. It is easy to check that $\left(\mc{P}^{\bk}\right)^2 = \mc{P}^{\bk}$, all $\mc{P}^{\bk}$ mutually commute, and that 
\beq
g(\mathbf{n}) \mc{P}^{\bk} = (-1)^{\bk \cdot \mathbf{n}} \mc{P}^{\bk}.
\eeq
This implies that $\{\mc{P}^{\bk}\}$ are the projection operators onto the simultaneous eigenstates of all of $G$ and, since $H_{\mc S}$ is a sum over a (over)complete generating set for $G$, onto the energy eigenvalue subspaces as labelled by $\bk$. Clearly, $\bk$ are the quantum numbers labeling excitations, where $\bk=0$ labels the ground state manifold.

The projectors $\mc{P}^{\bk}$ are pure state projectors if and only if $d_G = N$. In most cases under consideration, $d_G < N$ if the system is on a topologically non-trivial manifold, reflecting that all $\bk$ eigenstate manifolds---including the ground state manifold---of the Hamiltonian~\eqref{eq:stab} are degenerate on that manifold\footnote{This is a defining feature of systems with long-range entanglement; for SPTs, on the other hand, there is a unique ground state on arbitrary manifolds and, correspondingly, $d_G = N$ for those systems.}. To see this, one can take the trace of $\mc P^{\bk}$ and use the fact that tr$(g(\mathbf{n})) = N \delta_{\mathbf{n},0}$ to find that the dimension of the $\bk$ eigenvalue manifold is $2^{d_l}$ for $d_l= N-d_G$. We refer to this type of degeneracy as a {\it topological degeneracy}. To obtain pure state projectors, we can complete $G$ by adding $d_l$ Pauli operators to our generating set, such that these operators are mutually commuting and also commute with all elements $g \in G$. These $d_l$ operators, which preserve the ground state manifold (also referred to as the code-space in the context of error correction) but act non-trivially on it, are referred to as the \textit{logical operators} of the stabilizer code, a reflection of the fact that a stabilizer code encodes $d_l$ logical qubits in the ground space. Formally, the set of logical operators of a code is specified  by $L \coloneqq \mc{C}(G)/ G $, where $\mc{C}(G)$ is the centralizer of $G$.

%Since the Hamiltonian is simply a sum over group elements, the operators $\mc{P}^{(\bk)}$ project onto the various eigenvalue subspaces labelled by the quantum number $\bk$. This clearly labels excitations of the stabilizer code, with $\bk = 0$ the ground space. The dimension of these subspaces can be found by taking the trace and using the fact that tr$(g(\mathbf{n})) = N \delta_{\mathbf{n},0}$, leading to a \textit{topological} degeneracy $d_l = N - d_G$ for any $\bk$, as expected. \AS{I think we can remove this paragraph and include this info in the paragraph above.} 

In this paper, we study phases of matter captured by stabilizer code Hamiltonians of the form 
\beq
\label{eq:ham}
H = H_{\mc{S}} + \lambda V,
\eeq
where $H_{\mc{S}}$ is given by Eq.~\eqref{eq:stab} with $J_s = 1$ for all $s$ and where
\beq
\label{eq:pert}
V= \sum_q\sum_{i=1}^3 \xi^i_q X^i_q
\eeq
describes a perturbation to $H_\mc{S}$. Here, $\lambda$ is a control parameter, $\xi^i_q\in [-1,1]$ is an arbitrary real number, and $X^i_q$ is the $i^{th}$ single-qubit Pauli operator for the qubit $q$ ($1=X$, $2=Y$, and $3=Z$). We assume all $X_q^i$ anti-commute with at least one member of $\mc{S}$. We note that one could replace $X_q^i$ with any anti-commuting set of Pauli operators with local support (we are implicitly defining local as any Pauli operator with support less than that of any stabilizer). We pick all single-qubit operators for simplicity, although our results generalise straightforwardly. We also note that a large class of stabilizer code Hamiltonians are CSS codes, for which the Hamiltonian schematically takes the  form
\beq
\label{eq:CSS}
H_{\text{CSS}} = -\sum O_s^X - \sum O_s^Z,
\eeq
where $O_s^X$ ($O_s^Z$) is a product of only $X$-type ($Z$-type) Pauli operators. While all models considered in this paper are CSS codes, our results do not rely on this assumption. In what follows, we first show that in the absence of any perturbations, the entanglement spectrum of a stabilizer code Hamiltonian $H_{\mc{S}}$ is flat, and then proceed to study non-trivial universal features which appear in the ES by considering $V$ as a perturbation to $H_S$.

\subsection{Eigenstates and Entanglement Spectrum of Unperturbed Stabilizer Codes}
\label{sec:unpert}

Let us first consider a stabilizer Hamiltonian in the absence of any perturbations. As mentioned in the previous section, %for certain models, the order of the stabilizer group $d_G < N$ when the system is placed on a topologically non-trivial manifold \red{[cite]}. However, upon inclusion of $d_l = N - d_G$ logical operators,
we can extend $G \to \tilde G$ in Eq.~\eqref{eq:proj}, for $\tilde G$ generated by $\{O_i\}_{i\leq d_G}$ along with a mutually commuting set of logical operators. Then, $\bk \in \{0,1\}^N$ and 
\beq
\mc{P}^{\bk} = \outerp{\bk}{\bk},
\eeq
%such that $\{ \ket{\bk}\}$ forms the eigenbasis for the unperturbed Hamiltonian, hence allowing us to obtain pure state projectors and the density matrix.  
i.e. $\mc P^{\bk}$ is the pure state density matrix. For the energy of the state $\ket{\bk}$, when $|\mc S| \neq N$, then the terms $O_i$ from our generating set for $G$ in the Hamiltonian each contribute $-(-1)^{k_i}$, leading to a total energy 
\beq
E_\bk =  2\|\bk\| - d_G,
\eeq
where $\|\|$ is the Hamming weight. The contributions from the remaining terms ($\mc S-\{O_i\}_{i \leq d_G}$) depend on the specific generating set chosen but in general, these contributions will cancel in the energy denominators used in the perturbative expansion (see Sec.~\ref{sec:pert}). 

%The bits of $k$ which correspond to the logical operators are important here as they index the degeneracy which is associated with the ground space. Let $k_\ell, j_\ell \in \{0,1\}^N$ be any indices which correspond to ground states. 
%

We now calculate the entanglement spectrum of any eigenstate of the stabilizer Hamiltonian for a bipartition $(A,B)$, where $A$ is some subset of the qubits forming $\Hi$ and $B= A^c$. Following \cite{Haah2013,han2,albert}, one can show that for any $(A,B)$ such that $A$ is smaller than the code distance\footnote{The code distance is the minimum size of the support over all members of $L$.}, the reduced density matrix for any state $\bk$ is
 \begin{align} \label
 {eq:redDM}
 \rho_A = \text{Tr}_B \outerp{\bk}{\bk} =  2^{-s_A} \mc{P}^{\bk_A}_A,
 \end{align}
 where $s_A = |A|-d_{G_A}$ and $d_{G_A}$ is the dimension of the subgroup $G_A\subseteq G$ which only has support in $A$. $\mc{P}_A^{\bk_A}$ is the analogous projection operator to Eq.~\eqref{eq:proj} for $G_A$ and $\bk_A$ is the part of $\bk$ which corresponds to $G_A$.
 
 We have thus shown that the reduced density matrix is proportional to a projection operator, thereby proving that the ES of \textit{any} unperturbed stabilizer code is \textit{flat} \textit{i.e.,} eigenvalues of the reduced density matrix are all equal; in the process, we have also obtained the von Neumann entropy for $A$:
\beq
\label{eq:entent}
s_A = |A| - d_{G_A}.
\eeq
The same calculation can be carried out from the perspective of $B$ and we would find a nearly identical form of the reduced density matrix $\rho_B$ for the analogous $G_B$. Correspondingly, one can show that the entanglement entropy for $B$ is
\beq
s_B = |B| - d_{G_B} - d_l = s_A.
\eeq
%where we have used the fact that $N = d_G + d_l = |A| + |B|$.
These results also serve to highlight the relative simplicity and power of the stabilizer formalism, as we have made no assumptions besides a stabilizer description for the system of interest in the derivation. In order to extract interesting physics, one needs to further examine the behaviour of Eqs.~\eqref{eq:redDM} and~\eqref{eq:entent} in the presence of perturbations/deformations: for topologically ordered states, one is typically interested in extracting the sub-leading corrections to Eq.~\eqref{eq:entent} which are invariant under arbitrary deformations of the partition, while for SPT states, one expects a non-trivial degeneracy in the ES which is robust against arbitrary local perturbations respecting the symmetry protecting the bulk state. We relegate the discussion of such non-trivial features to Secs.~\ref{sec:edge-es} and~\ref{sec:sspt}, where we analyse the ES of various models in the presence of perturbations. 

A related concept to the entanglement entropy for stabilizer codes is that of the \textit{recoverable information}, defined for any stabilizer code and bipartition $(A,B)$ as
\beq
\label{eq:recover}
\mu= \min( d_\partial -s_A -s_B),
\eeq
where $d_\partial$ is the number of members of $\{O_i\}_{i\leq d_G}$ which are ``cut,'' \textit{i.e.,} have support in both $A$ and $B$, and the minimization is over all possible choices of a generating set, assuming open boundary conditions. In Ref.~\cite{albert}, it was shown that $\mu>0$ and, generally, that $d_{\partial} -s_A -s_B$ is equal to the dimension of the {\it non-local surface stabilizer} (NLSS) group $G_{\text{NLSS}}$ defined as 
\beq
G_{\text{NLSS}} = G_\partial \cap (G_A \oplus G_B),
\eeq
where $G_\partial$ is the group generated by the cut members of $\{O_i\}_{i\leq d_G}$. Thus, the recoverable information is the minimum dimension over all NLSS groups. Every member of $g_\partial \in G_{\text{NLSS}}$, referred to as an NLSS, has the general form 
\beq
g_\partial= g_A g_B,
\eeq
where $g_A \in G_A$ and $g_B \in G_B$. We can interpret each of these as an \textit{emergent Gauss' law} constraint by noting that if $g_A = \prod_{i \in F_A} O^A_i$, for stabilizers $O_i^A \in G_A$ and $F_A$ a subset of stabilizer indices in $A$, then by restricting the NLSS to its support in $A$ (where we note $(g_B)_A =I_A$) we have
\beq\label{eq:gauss}
\prod_{i \in F_A} O^A_i= (g_\partial)_A.
\eeq
Thus, a measurement purely on the boundary is equal to the number of ``charges'' in a subset of the bulk (mod 2). In general, either $g_A$ or $g_B$ has the given form, but it may happen that $g_A = I_A$ and $g_B \neq I_B$, or vice versa, \textit{i.e.,} 
\beq
g_\partial = g_B,
\eeq
or some product of cut stabilizers is only supported in $B$. We refer to such an NLSS as a {\it superficial NLSS} since the corresponding Gauss' law is \textit{only} along the boundary. Every constraint which contains a cut stabilizer always implies an NLSS. However, not all NLSS are formed this way. The minimization in the definition of recoverable information removes all NLSS coming from trivial constraints with the aim of capturing only those arising from topological constraints. %\SJ{[This is not obvious to me. Is it a result in Albert's previous paper?]}\AS{Yes.}
We return to the importance of NLSS and how they protect the emergent Gauss' laws in the ES in Sec.~\ref{sec:summary}.

Before proceeding to the analysis of the ES of stabilizer codes in the presence of perturbations, we need to establish some further properties of the eigenstates $\ket{\bk}$, for which we can infer a Schmidt decomposition for a given cut $(A,B)$ from Eq.~\eqref{eq:redDM}. Any stabilizer in $A$, $O^A_i \in G_A$, maintains the same eigenstate with respect to the reduced density matrix $\rho_A$, thus allowing us to partially index all Schmidt vectors by their eigenvalues for the stabilizers in $A$. We then require a basis for the subspace defined by the projectors $\mc{P}_A^{\bk_A}$, and we use cut stabilizers for this purpose. Likewise, we do the same for $G_B$  %\AP{May be useful to have an illustrative figure.}

Since all of $G_A$ is supported in $A$, the support of any cut stabilizer in $A$ necessarily commutes with all of $G_A$. Further, as the recoverable information is always positive, there exist more cut stabilizers than $\log_2$ of the rank of $\mc{P}^{\bk_A}_A$. Thus, it is always possible to choose a mutually commuting set of cut stabilizers (with no unique choice for this set) whose simultaneous eigenstates span this space and (nearly) suffice to completely label the Schmidt vectors. Let $G^R_\partial \subseteq G_\partial$ be the reduced boundary group generated by the chosen basis for the cut stabilizers. Members of $G^R_\partial$ can be indexed by $\bk_{\partial} \in \{0,1\}^s$, where $s = s_A = s_B$ is the entanglement entropy. Then the Schmidt decomposition can be written as
\begin{align}\label{eq:schmidt}
\ket{\bk}= 2^{-\frac{s}{2}} \sum_{\mbf{l}_{\partial}} (-1)^{\mbf{l}_\partial \mbf{M} \bk} \ket{\bk_A, \mbf{l}_\partial}_A \otimes \ket{\bk_B, \mbf{l}_\partial \oplus \bk_\partial}_B,
\end{align}
with
\beq
(O_i)_A\ket{\bk_A, \mbf{l}_\partial}_A = (-1)^{(\mbf{l}_{\partial})_i}\ket{\bk_A, \mbf{l}_\partial}_A,
\eeq
where $O_i$ is a generator of $G^R_\partial$, and likewise for $B$. Note that the Schmidt decomposition Eq.~\eqref{eq:schmidt} is not unique as a result of the flat ES; the choice made here is informed by the fact that $\ket{\bk}$ must have the same stabilizer eigenvalues in its Schmidt form. This is why the two boundary indices in each term must have a binary sum of $\bk_\partial$. Likewise, $\mbf{M}$ is an $s \times d_G$ binary matrix which maintains the eigenvalues for the stabilizers in $G_\partial - G^R_\partial$. Such operators must anti-commute with both pieces of some cut stabilizers in $G_\partial$ (since it commutes with the complete stabilizers). This flips the $\mbf{l}_\partial$ eigenstates by some $\mbf{p}^i \in \{0,1\}^s$. Upon relabelling the sum, we find that $\mbf{M}$ must satisfy $\mbf{p}^i \mbf{M} \bk = \bk_i$. 

\subsection{Perturbed Density Matrix and Entanglement Hamiltonian}
\label{sec:pert}

Thus far, we have shown that, in the absence of any perturbations, the reduced density matrix for a stabilizer Hamiltonian~\eqref{eq:stab} is proportional to a projection operator; consequently, the unperturbed entanglement spectrum is flat. We now consider the Hamiltonian~\eqref{eq:ham}, which includes perturbations of the form specified in Eq.~\eqref{eq:pert}, to identify universal features of the ES which persist in the presence of such perturbations. Before delving into the full derivation, we outline the general idea underlying our method for finding an approximate density matrix for the perturbed ground state \textit{i.e.}, the ground state of Eq.~\eqref{eq:ham}. We make the following general assumptions:
\begin{itemize}
\item The perturbation is weak enough so as to not close the energy gap \textit{i.e.,} we remain in the same phase described by the exactly solvable stabilizer Hamiltonian. 
\item We work in the thermodynamic limit of the system.
\item We require that the sub-region $A$, although much smaller than the entirety of the system, is much larger than the size of the stabilizers (region over which a stabilizer has non-trivial support) \textit{i.e.,} the linear size $R$ of region $A$ is assumed to obey $\xi \ll R \ll L$, where $\xi$ is the correlation length of the perturbed Hamiltonian and $L$ is the linear size of the system.
\end{itemize}

We make use of unitary perturbation theory (see Appendix~\ref{UPT}), which is a variation on the oft-utilised Schrieffer-Wolff perturbation theory~\cite{swpert,bravyisw}. The central focus of unitary perturbation theory (UPT) is to approximate a unitary operator $U$ which maps unperturbed eigenvectors $\ket{n}$ to eigenvectors $\ket{n'}$ of the perturbed system. As with any controlled perturbative scheme, this can be done up to some fixed order in the perturbative parameter $\lambda$. Where UPT differs from conventional perturbation theory is that while the latter truncates the expansion of the system state $\ket{\psi}$ to a given order in $\lambda$, the former instead truncates the expansion of the anti-hermitian generator of the unitary $U$ to that order in $\lambda$. This ensures that the approximate transformation $U$ maintains unitarity at any finite order in its perturbative expansion. To first order in the control parameter $\lambda$, the unitary operator is given by $U = \exp(\lambda \mc{L})$, with
\begin{align} \label{eq:Lstab}
\mc{L} = - \sum_{\bk,\mbf{l}} \frac{V_{\bk\mbf{l}} \large[E_\bk\neq E_\mbf{l}\large]}{E_\bk- E_\mbf{l} + 0} \outerp{\bk}{\mbf{l}},
\end{align}
and where $V_{\bk \mbf{l}} = \braket{\bk|V|\mbf{l}}$ and $[.]$ is the Iverson bracket which equals 1 if the proposition inside is true and 0 otherwise (see Appendix~\ref{UPT}). Making a further approximation valid in the large $A$ limit, we consider the resulting action of the unitary $U$ on the Schmidt form Eq.~\eqref{eq:schmidt}, allowing us to perform the partial trace required to form the perturbed density matrix $\tilde \rho_A$ for the ground state of the perturbed Hamiltonian. The perturbed entanglement Hamiltonian (EH), to first order, is then defined as
\begin{align}
H_{\text{ent}} \simeq -\left(\prtl{\tilde \rho_A}{\lambda}\right)_{\lambda=0},
\end{align}
where $\simeq$ implies unitary equivalence. Hence, $\lambda$ may be thought of as the inverse temperature for this state.  

We now describe the derivation of $H_{\text{ent}}$ in detail, starting with the matrix elements of the perturbation, Eq.~\eqref{eq:pert}. Recall that $X_q^i$ is the $i^{th}$ single-qubit Pauli operator for qubit $q$, with $1=X, 2=Y,$ and $3=Z$. For every $X_q^i$, we can assign a binary string $\mbf{p}_q^i \in \{0,1\}^{N}$ representing all stabilizer basis elements which anti-commute with $X_q^i$. Thus, 
\beq
P^\bk X_q^i = X_q^i P^{\bk\oplus \mbf{p}_q^i},
\eeq
where $\oplus$ is the binary sum or bitwise-XOR of the two strings. From this we get the modulus-square of each perturbation term
\begin{align}
|\braket{\bk|X_q^i|\mbf{l}}|^2 = \braket{\bk | X_q^i P^\mbf{l} X_q^i | \bk} = \delta_{\bk \oplus \mbf{l}, \mbf{p}_q^i}.
\end{align}
However, this does not resolve the phase of the matrix element: $ X_q^i \ket{\bk} \propto \ket{\bk \oplus \mbf{p}_q^i}$, up to an overall phase. Nonetheless, one can complete a canonical basis for all Pauli operators as described in Appendix~\ref{ap:canon} using the stabilizer (and logical) basis operators and some canonical duals. This allows us to write all $X_q^i$ in terms of this basis, which implies the existence of a $\tilde{\mbf{p}}_q^i \in \{0,1\}^{N}$ such that
\begin{align}
\braket{\bk|X_q^i|\mbf{l}}= (-1)^{\bk \oplus \tilde{\mbf{p}}_q^i} \, \delta_{\bk \oplus \mbf{l}, \mbf{p}_q^i}.
\end{align}
Thus, the matrix element $V_{\bk \mbf{l}}$ is given by
\beq
V_{\bk \mbf{l}} = \braket{\bk|V|\mbf{l}} = \sum_q \sum_{i=1}^3 (-1)^{\bk \oplus \tilde{\mbf{p}}_q^i} \, \xi_q^i \, \delta_{\bk \oplus \mbf{l}, \mbf{p}_q^i}.
\eeq
We can also rewrite the energy denominator as
\begin{align}
E_\bk -E_{\bk \oplus \mbf{p}_q^i} = 2 (\|\bk\oplus \mbf{p}_q^i \| - \|\bk\|)= 2(\overline \bk \cdot \mbf{p}_q^i -  \bk \cdot \mbf{p}_q^i), 
\end{align}
where the overbar denotes the complement, or negation, of the string.

Putting the terms together, we see that the generator of the perturbation unitary is  
\begin{align}
\mc{L} = \sum_\bk \sum_q \sum_{i=1}^3  C_\bk^{q_i} \outerp{\bk}{\bk \oplus \mbf{p}_q^{i}},
\end{align}
where we have defined the $C$-coefficient
\begin{align}\label{eq:coef1}
C_\bk^{q_i} = \frac{1}{2} \frac{(-1)^{\bk \cdot \tilde{\mbf{p}}_q^{i}} \, \xi_q^{i} \, [\bk\cdot \mbf{p}_q^{i} \neq \bar{\bk}\cdot  \mbf{p}_q^{i}] }{\overline \bk \cdot \mbf{p}_q^i -  \bk \cdot \mbf{p}_q^i+ 0}.
\end{align}
Despite the apparent complexity of the $C$-coefficients, a crucial property which can be easily established is that for qubits $q, r$ which are not both contained within the support of some stabilizer, 
\beq
\label{eq:Cprop}
C_{\bk \oplus \mbf{p}_r^j}^{q_i} = C_{\bk}^{q_i},
\eeq
which reflects the fact that if two qubits are far enough separated, they do not interact. This becomes clear upon examination of the energy denominator, which is sensitive only to the string in the vicinity of $q$. In other words, the binary dot product projects $\bk$ and $\bk \oplus \mbf{p}_r^j$ onto $\mbf{p}_q^i$ and, if the two qubits $q,r$ are not both contained within the support of some stabilizer (are far enough away), the projection is not affected by the change $\bk \to \bk \oplus \mbf{p}_r^j$. As for the overall phase, the invariance of the $C$-coefficients under this change can be seen from 
\beq
(-1)^{(\bk \oplus \mbf{p}_r^j) \cdot \tilde{\mbf{p}}_q^i} = (-1)^{\bk \cdot \tilde{\mbf{p}}_q^i} \, (-1)^{\mbf{p}_r^j \cdot \tilde{\mbf{p}}_q^i} = (-1)^{\bk \cdot \tilde{\mbf{p}}_q^i},
\eeq
where we have used the (anti) commutation relations for the strings under the assumption $q \neq r$. Another important property of the $C$-coefficients can be established from the requirement that $\mc{L}$ is skew-Hermitian: 
\beq
C^{q_i}_{\bk \oplus \mbf{p}_q^i} = - C^{q_i}_\bk.
\eeq

Expressing the skew-Hermitian generator of the perturbation unitary in terms of the $C$-coefficients and using their aforementioned properties allows us to expand the perturbation unitary $U$ as 
\begin{align}
U = \exp(\lambda \mc{L}) = \sum_\alpha \frac{\lambda^\alpha}{\alpha!}\sum_\bk \sum_{\vec{q} \in Q^{\otimes \alpha}} C_{\bk \oplus \mbf{p}_{\vec{q}}}^{\vec{q}} \outerp{\bk}{\bk \oplus \mbf{p}_{\vec{q}}}, \label{eq:powers}
\end{align}
where, for ease of notation, we have suppressed the Pauli-type index, and defined $\mbf{p}_{\vec{q}} \equiv \oplus_i \mbf{p}_{q_i}$ and
 \begin{align}
 \label{eq:cvec}
C_{\bk \oplus \mbf{p}_{\vec{q}}}^{\vec{q}}= C_{\bk}^{q_1} C_{\bk \oplus \mbf{p}_{q_1}}^{q_2}C_{\bk \oplus \mbf{p}_{q_1} \oplus \mbf{p}_{q_2}}^{q_3} \dots C_{\bk \bigoplus_i^{\alpha-1}\mbf{p}_{q_i}}^{q_\alpha}.
\end{align}
 Eq.~\eqref{eq:powers} can be verified through induction.
 
 In order to perform the partial trace, we make a further approximation that for any string $\vec{q}$, we can reasonably make the substitution, 
 \begin{align}
 \label{eq:approx}
 C_{\bk \oplus \mbf{p}_{\vec{q}}}^{\vec{q}} \to C_{\bk \oplus \mbf{p}_{\vec{q}_A}}^{\vec{q}_A} C_{\bk \oplus \mbf{p}_{\vec{q}_B}}^{\vec{q}_B},
 \end{align}
 where $\vec{q}_{A(B)}$ is the part of the string $\vec{q}$ contained in region $A(B)$ while maintaining their relative order. While the binary sum in the bra, $ \mbf{p}_{\vec q} =\oplus_i \mbf{p}_{q_i} =\mbf{p}_{\vec q_A} \oplus \mbf{p}_{\vec q_B}$ is exact, the corresponding coefficient in the expansion of the perturbation unitary $U$ is not, making this a non-trivial approximation whose effect must be closely evaluated. From the definition of the $C$-coefficients Eq.~\eqref{eq:coef1}, the fact that $\xi_q \in [-1,1]$, and that the energy gap is always bounded by $1$, it is straightforward to see that
 \begin{align}
 \label{eq:bound}
 \left| C_{\bk \oplus \mbf{p}_{\vec{q}}}^{\vec{q}} - C_{\bk \oplus \mbf{p}_{\vec{q}_A}}^{\vec{q}_A} C_{\bk \oplus \mbf{p}_{\vec{q}_B}}^{\vec{q}_B}\right| \leq 1.
 \end{align}
 Hence, to evaluate the effect of the approximation Eq.~\eqref{eq:approx}, we must consider how many terms in the sum are in error. It should be clear from the definition of the $C$-coefficients that only the energy denominators need to be considered; since the energy denominators ``project'' on the $\bk$ string in the vicinity of $q$, if $q_1$ and $q_2$ are not shared within the support of any stabilizer, Eq.~\eqref{eq:Cprop} implies that $C^{q_2}_{\bk\oplus \mbf{p}_{q_1}} = C^{q_2}_\bk$. At first order, our approximation is clearly exact. At second order, suppose that we are considering a $d$-dimensional system such that the size of the system $\sim L^d$ for $L$ large and that the size of region $A \sim R^d$ with $R <L$. Of the $\sim L^{2d}$ total terms, only $\sim R^{d-1}$ are in error as only this number of terms have both qubits in the boundary and in the same cut stabilizer. The number of terms in error thus comprise a vanishing, negligible fraction in the thermodynamic limit. On the other hand, one might worry that only terms which involve the boundary will matter. However, even considering only those terms for which both qubits are in the boundary, the ratio of error terms to non-error terms goes as $R^{-(d-1)}$, and so the error is suppressed as $R$ becomes large (assuming $d\geq 2$). The suppression of higher-order errors follows by a similar logic.
 
Effectively, the replacement~\eqref{eq:approx} amounts to the assumption that all terms for which $\vec{q}$ is composed of $\vec{q}_A$ and $\vec{q}_B$ are equal. We must thus account for the multiplicity of these terms. For $\vec{q}_A$ of length $\beta$ and $ \vec{q}_B$ of length $\alpha -\beta$, the multiplicity is $ \binom{\alpha}{\beta}$.

We now apply the perturbation unitary $U$ to a ground state $\ket{0}$ of the unperturbed Hamiltonian in order to find the approximate ground state for the perturbed Hamiltonian~\eqref{eq:ham}. Using the Schmidt form of the unperturbed eigenvectors Eq.~\eqref{eq:schmidt}, and within the approximation~\eqref{eq:approx}, we find that
\begin{widetext}
\begin{align}
\label{eq:split}
U\ket{0} &= \sum_\alpha \frac{(-\lambda)^\alpha}{\alpha!} \sum_{\vec{q} \in Q^{\otimes \alpha}} C_{ \mbf{p}_{\vec{q}}}^{\vec{q}} \ket{\mbf{p}_{\vec{q}}} \nonumber \\
&\approx \sum_\alpha \frac{(-\lambda)^\alpha}{\alpha!} \sum_{\vec{q}_A \in A^{\otimes \alpha}} C_{\mbf{p}_{\vec{q}_A}}^{\vec{q}_A} \sum_\beta \frac{(-\lambda)^\beta}{\beta!} \sum_{\vec{q}_B \in B^{\otimes \beta}} C_{\mbf{p}_{\vec{q}_B}}^{\vec{q}_B} \ket{\mbf{p}_{\vec{q}_A} \oplus \mbf{p}_{\vec{q_B}}} \nonumber \\
&= 2^{-\frac{s}{2}}\sum_{\bk_\partial} \left( \sum_\alpha \frac{(-\lambda)^\alpha}{\alpha!} \sum_{\vec{q}_A \in A^{\otimes \alpha}}(-1)^{\bk_{\partial} \mbf{M} \mbf{p}_{\vec{q}_A}} C_{\mbf{p}_{\vec{q}_A}}^{\vec{q}_A} \ket{\left(\mbf{p}_{\vec{q}_A} \right)_{A\cup \partial A}\oplus \mbf{k}_\partial}_A \right) \nonumber \\
&\otimes
\left( \sum_\beta  \frac{(-\lambda)^\beta}{\beta!}\sum_{\vec{q}_B \in A^{\otimes \beta}}(-1)^{\bk_{\partial} \mbf{M} \mbf{p}_{\vec{q}_B}} C_{\mbf{p}_{\vec{q}_B}}^{\vec{q}_B} \ket{\left(\mbf{p}_{\vec{q}_B} \right)_{B\cup \partial B}\oplus \bk_\partial}_B \right) \nonumber \\
&:=  2^{-\frac{s}{2}} \sum_{\bk_\partial} \ket{\psi_{\bk_\partial}}_A \otimes \ket{\phi_{\bk_\partial}}_B,
\end{align}
\end{widetext}
where by $\left(\mbf{p}_{\vec{q}_A} \right)_{A\cup \partial A}$ we mean the the part of $p_{\vec{q}_A}$ corresponding to $G_A$ and $G_\partial^R$, and likewise for $B$. $\bk_\partial$ is short for $(\mbf{0}_A, \bk_\partial)$, which labels members of $G_A \oplus G^R_\partial$. Note in particular that we have split-up the binary values of $\mbf{p}_{\vec{q}_A} \oplus \mbf{p}_{\vec{q_B}}$ such that if $q \in A$, then it only contributes to $G_A \oplus G^R_\partial$ in the Hilbert space of $A$. This is always possible since an edge in $A$ can only anti-commute with members of $G_A \oplus G^R_\partial$. Likewise is true for $B$. This splitting ensures that the binary sum of the values in the $A$ and $B$ kets equals the value in the original ket, as per our discussion in Section~\ref{sec:unpert}. 

While the final expression obtained in Eq.~\eqref{eq:split} is ostensibly another Schmidt form for the approximate perturbed ground state, this assumes that $\{\ket{\psi_{\bk_\partial}}_A\}$ and $\{ \ket{\phi_{\bk_\partial}}_B\}$ are orthonormal. Focusing on $A$, we consider the overlap of any two of these states,
\begin{widetext}
\begin{align}
\braket{\psi_{\bk_\partial}|\psi_{\mbf{l}_\partial}} =\sum_{\alpha, \beta} \frac{(-\lambda)^{\alpha+\beta}}{\alpha!\beta!} \sum_{\vec{q}_1 \in A^{\otimes \alpha}}\sum_{\vec{q}_2\in A^{\otimes \beta}}(-1)^{\bk_{\partial} \mbf{M} \mbf{p}_{\vec{q}_1}}(-1)^{\mbf{l}_{\partial} \mbf{M} \mbf{p}_{\vec{q}_2}} C_{\mbf{p}_{\vec{q}_1}}^{\vec{q}_1}C_{\mbf{p}_{\vec{q}_2}}^{\vec{q}_2}  \braket{\left(\mbf{p}_{\vec{q}_1} \right)_{A\cup \partial A}\oplus \bk_\partial|\left(\mbf{p}_{\vec{q}_2} \right)_{A\cup \partial A}\oplus \mbf{l}_\partial}.
\end{align}
\end{widetext}
If $\bk_\partial = \mbf{l}_\partial$, then $\left(\mbf{p}_{\vec{q}_1} \right)_{A\cup \partial A} =\left(\mbf{p}_{\vec{q}_2} \right)_{A\cup \partial A}$ within the sum, causing the phase factors to cancel and all dependence on $\bk_\partial$ drops out. This implies that $\braket{\psi_{\bk_\partial}|\psi_{\bk_\partial}}$ is a constant which, without loss of generality, we take to be $1$. However, when $k_\partial \neq l_\partial$, the fact that the $C$-coefficients are insensitive to $\bk_\partial, \mbf{l}_\partial$ implies that the inner product does not sum to zero. Nonetheless, we can still make use of this form to perform the partial trace, which is given by
\begin{align}
\tilde \rho_A &\approx \frac{1}{\mc Z} \sum_{\bk_\partial,\mbf{l}_\partial} \braket{\phi_{\mbf{l}_\partial}|\phi_{\bk_\partial}} \outerp{\psi_{\bk_\partial}}{\psi_{\mbf{l}_\partial}} \nonumber \\
&= \frac{1}{\mc{Z}} \sum_{\bk_\partial} \outerp{\psi_{\bk_\partial}}{\psi_{\bk_\partial}} + \frac{1}{\mc Z}\sum_{\bk_\partial \neq \mbf{l}_\partial} \braket{\phi_{\mbf{l}_\partial}|\phi_{\bk_\partial}} \outerp{\psi_{\bk_\partial}}{\psi_{\mbf{l}_\partial}},
\end{align}
where $\mc{Z}$ is the normalisation factor which makes $\tilde \rho_A$ have unital trace after the approximation is made. To further simplify the expression, we define
\begin{align}
\tilde U_A &= \sum_\alpha \frac{(-\lambda)^\alpha}{\alpha!} \sum_{\bk_A, \bk_\partial} \sum_{\vec{q} \in A^{\otimes \alpha}} \Bigg( (-1)^{\bk_\partial \mbf{M} \mbf{p}_{\vec{q}}} \, C_{(\bk_A, 0) \oplus \mbf{p}_{\vec{q}}}^{\vec{q}} \nonumber \\
& \times \outerp{(\bk_A, \bk_\partial) \oplus \mbf{p}_{\vec{q}}}{(\bk_A, \bk_\partial)} \Bigg) 
\end{align}
and analogously for $\tilde U_B$. Using these definitions, we can write 
\beq
\ket{\psi_{\bk_\partial}} = \tilde U_A \ket{\bk_\partial}, 
\eeq
and
 \begin{align}
 \label{eq:reddinter}
 \mc{Z} \tilde \rho_A &\approx \tilde U_A\left( \mc{P}_A^\mbf{0}  + \sigma_B \right) \tilde U_A^\dagger \nonumber \\
 & \simeq \sqrt{ \mc{P}_A^\mbf{0}  + \sigma_B} \left( \tilde U_A^\dagger  \tilde U_A\right) \sqrt{ \mc{P}_A^\mbf{0}  + \sigma_B},
 \end{align}
where
\begin{align}
\sigma_B &= \sum_{\bk_\partial\neq \mbf{l}_\partial} \braket{\phi_{\mbf{l}_\partial}|\phi_{\bk_\partial}} \outerp{\bk_\partial}{\mbf{l}_\partial} \nonumber \\
&= \sum_{\bk_\partial\neq \mbf{l}_\partial} \left( \mc{P}_B^\mbf{0}\tilde U_B^\dagger \tilde U_B \mc{P}_B^\mbf{0}\right)_{\mbf{l}_\partial \bk_\partial} \outerp{\bk_\partial}{\mbf{l}_\partial} \in \mc O(\lambda),
\end{align}
with $\mc{P}_A^\mbf{0}$ ($\mc{P}_B^\mbf{0}$) a projector for $G_A (G_B)$ (see Sec.~\ref{sec:unpert}). 

Note that unitary equivalence in Eq.~\eqref{eq:reddinter} is a result of the fact that, for two operators $O_1$ and $O_2$, $O_1 O_2$ has the same non-zero eigenvalues as $O_2 O_1$. Since we are only interested in the density matrix up to first order in the control parameter $\lambda$ and since $\sigma_B \in \mc O(\lambda)$, we can expand the square root as $\sqrt{ \mc{P}_A^\mbf{0}  + \sigma_B} = \mc{P}_A^\mbf{0}  + \hlf\sigma_B + \mc O(\lambda^2)$. Again, to lowest order we find that
\begin{align}
\mc{Z} \tilde \rho_A \approx\mc{P}_A^\mbf{0}\tilde U_A^\dagger  \tilde U_A \mc{P}_A^\mbf{0} + \sigma_B,
\end{align}
where we have made use of the fact that $\tilde U_A$ is the identity at lowest order and that $\mc{P}_A^\mbf{0} \sigma_B = \sigma_B\mc{P}_A^\mbf{0} = \sigma_B$.

We are now well positioned to define the two parts of the EH:
\begin{subequations}
\begin{align}
\label{eq:HentA}
H_{\text{ent}}^A=& - \left(\prtl{\mc{P}_A^\mbf{0} \tilde U_A^\dagger  \tilde U_A \mc{P}_A^\mbf{0}}{\lambda}\right)_{\lambda=0}, \\
H_{\text{ent}}^B=& -\left(\prtl{\sigma_B}{\lambda}\right)_{\lambda=0} \nonumber \\
=& \sum_{\bk_\partial\neq \mbf{l}_\partial} \left(\prtl{\left( \mc{P}_B^\mbf{0} \tilde U_B^\dagger \tilde U_B\mc{P}_B^\mbf{0}\right)_{\mbf{l}_\partial \bk_\partial}}{\lambda}\right)_{\lambda=0} \outerp{\bk_\partial}{\mbf{l}_\partial}.
\end{align}
\end{subequations}
The $A, B$ symmetry manifest in these two terms should be unsurprising given that the entanglement spectrum is the same for both $A$ and $B$, with contributions naturally arising from both sides of the boundary. This also suggests that the non-triviality of the EH is a consequence of the non-unitarity of the operators $\tilde U_A$ and $\tilde U_B$. Evaluating the expression~\eqref{eq:HentA}, we find in general that
\begin{align}
\label{eq:EntHA}
H_{\text{ent}}^A &= 2 \sum_{\bk_\partial} \sum_{q \in A}(-1)^{\bk_\partial \mbf{M} \mbf{p}_q} C^q_{\mbf{p}_q} \mc{P}_A^\mbf{0}\outerp{\bk_\partial \oplus \mbf{p}_q}{\bk_\partial}\mc{P}_A^\mbf{0} \nonumber \\
&= \sum_{q\in \partial A} \frac{\xi_q}{\|\mbf{p}_q\|} \tilde X_q,
\end{align}
from which we immediately find that
\begin{align}
\label{eq:EntHB}
H_{\text{ent}}^B &= \sum_ {q \in \partial B} \frac{\xi_q}{\|\mbf{p}_q\|} \sum_{\bk_\partial\neq \mbf{l}_\partial} (-1)^{\bk_\partial \mbf{M} \mbf{p}_q}  \left(\tilde X_q\right)_{\mbf{l}_\partial \bk_\partial} \outerp{\bk_\partial}{\mbf{l}_\partial} \nonumber \\
&= \sum_ {q \in \partial B} \frac{\xi_q}{\|\mbf{p}_q\|} \tilde Z_q.
\end{align}
Equations~\eqref{eq:EntHA} and~\eqref{eq:EntHB} constitute the central results of this paper, upon which we will now elaborate. 

Let us consider $H_{\text{ent}}^A$~\eqref{eq:EntHA}, where for $q \in A$ $\tilde X_q = \mc{P}_A^\mbf{0} X_q\mc{P}_A^\mbf{0}$, such that the only perturbations which survive are those which commute with \textit{all of $G_A$}---hence, only members of $\partial A$ need be considered. Crucially, we note that there often exist \textit{constraints} on the terms of the (perturbative) EH: if any product of terms forms a member of $G_A$, \textit{i.e.,} there exists a subset $F_\partial \subseteq \partial A$ such that $\prod_{q \in F_\partial}X_q \in G_A$, the projection  must satisfy
\beq
\label{eq:constraint}
\prod_{q \in F_\partial }\tilde X_q = \mc{P}_A^\mbf{0}. 
\eeq
This constraint is a consequence of the NLSS, as discussed in Sec.~\ref{sec:unpert} where 
$
\prod_{q \in F_\partial }\tilde X_q= \left(g_\partial\right)_A = \prod_{i \in F_A} \mc O_i
$
for some set of indices $F_A$, as described in Eq.~\ref{eq:gauss}. Such a constraint indicates the presence of non-trivial %topological
 entanglement features in the model under consideration. Equivalently, the constraint~\eqref{eq:constraint} may be thought of as defining a $\mb Z_2$ topological surface charge, with the entanglement Hamiltonian confined to the zero charge sector; we return to a closer examination of these features in Sec.~\ref{sec:summary}. 

Let us now turn to $H_{\text{ent}}^B$~\eqref{eq:EntHB}, for which we must consider the form of the operators $\tilde Z_q$ for $q \in B$. If $X_q$ commutes with all of $G_B$, then its action must flip only cut stabilizers, implying that $\tilde Z_q$ is an operator which flips the same cut stabilizers, but with respect to their support in $A$. Such an operator can always be found by considering a canonical basis for Pauli operators in $A$ by using a set of $|A|$ complete and cut stabilizers in $A$, and then finding their canonical duals. Thus, $\tilde Z_q$ is necessarily the Pauli operator formed as the product of all operators dual to the cut stabilizers with which $\tilde Z_q$ anti-commutes. As constructing $\tilde Z_q$ in this manner tends to be a fairly tedious procedure in practice, we now discuss a simpler, more efficient method.

Since $\tilde Z_q$ is a Pauli operator, let us consider $ Z_q X_q$ (where the lack of a tilde signifies that we are considering the corresponding Pauli operator in the full space and without the projection). As $Z_q$ and $X_q$ anti-commute with the same members of $ \mc S$, their product necessarily commutes with all of $G$. This implies $Z_q X_q \in \tilde G$, \textit{i.e.,} it is either a product of stabilizers or a logical operator. If $A$ is significantly smaller than the code distance and $X_q$ is local, we can generally conclude that $Z_q X_q$ can not be a logical operator. Thus $Z_q X_q$ must be a member of $G_\partial$ such that its only support in $B$ is $X_q$ and only support in $A$ is $Z_q$ {\it i.e.} it represents a cut stabilizer group element. Further, just as was the case with $\tilde X_q$, all $\tilde Z_q$ are subject to the same constraint that some product of these operators must act as $\mc{P}_A^\mbf{0}$ if they form an NLSS. By definition, if there exists a set $F_\partial \subseteq \partial B$ such that $\prod_{q \in F_\partial}X_q$ forms an NLSS in $G_B$, $\prod_{q \in F_\partial}Z_q$ forms the corresponding NLSS in $G_A$ as every NLSS has the form $g_\partial = g_A g_B$ and $g_A$ is unique to $g_B$ (see Ref.~\cite{albert}). Note that the only guarantee that $Z_q$ is local is if $X_q Z_q \in \mc S$, \textit{i.e.}, the product forms a single stabilizer.

We note that even though we included only single-qubit Pauli operators in the perturbation $V$ to the stabilizer Hamiltonian, if all such terms of a given Pauli type do not survive the projection, we are then forced to consider higher order perturbations. Rather than computing second-order corrections coming from UPT, we can instead add local perturbations consisting of two-qubit Pauli operators. All of the results from above follow in a similar fashion, with the caveat that we now have to carefully consider those perturbation terms which are themselves cut. However, since every single-qubit Pauli does not survive, such cut perturbation terms will also not survive and can hence be safely ignored. We refer to such a process as {\it second-order} even though it technically arises at first-order in UPT, albeit from two-qubit perturbations\footnote{The use of this term is justified as, even for single-qubit Pauli operator perturbations, we expect that the EH contains higher-order terms of the same form but with different coefficients. In other words, second order UPT with single qubit Pauli operator perturbations yields the same results, up to unimportant coefficients, as first order UPT with two qubit Pauli operator perturbations. Our choice to proceed with the latter is made simply to avoid a lengthy digression into the derivation of second order UPT coefficients.}. Analogously, if all second-order contributions fail to survive, we continue to third-order contributions and so forth, until all lowest order contributions are identified.

\subsection{Entanglement Hamiltonian for different topological sectors} 

Until this point, we have effectively assumed the ground state under consideration is the $+1$ eigenstate under all logical operators. For $A$ smaller than the code distance, the cleaning lemma~\cite{bravyiterhal,haahreview} ensures that no logical operator is cut and thus its eigenstate does not affect the resulting entanglement Hamiltonian. However, if $A$ is a non-cleanable subset of the qubits, then some logical operators are necessarily cut. What this means is simply that some subset of the terms may not form NLSS's but rather form logical operators. If the original ground state is in the $-1$ eigenstate of some logical operator, then this has the effect of projecting the entanglement Hamiltonian onto the one-charge sector for the topological surface charge defined by that logical operator. Since both NLSS's and logical operators can be connected to topological constraints amongst the stabilizers (see Refs.~\cite{albert, albertgauge}), we can define these topological surface charges in a bipartition-independent way (\textit{i.e.,} a topological surface charge is defined by its relation to a topological constraint). This also has the rather interesting implication that a topological surface charge %can always be connected to (\AP{
can always measure a topological charge in the bulk, which constitutes a bulk-boundary correspondence for topologically ordered phases. We return to a discussion of topological surface charges and their relation to topological constraints in Sec.~\ref{sec:summary}.

%%%%%%%%%%%%%%%%%%%%%%%%%%%%%%%%%%%%%%%%

\section{Entanglement Spectra for Long-Range Entangled States}
\label{sec:edge-es}

Having established the general formalism for deriving the entanglement spectra for stabilizer codes in the presence of generic perturbations, we now analyse the resultant entanglement Hamiltonians given by Eqs.~\eqref{eq:EntHA} and~\eqref{eq:EntHB} for specific models. In this section, we discuss stabilizer codes describing LRE quantum phases of matter, which include the $d=2$ toric code/Wen-plaquette model and $d \geq 3$ toric code as examples of conventional topological order. We then consider the X-cube model~\cite{fracton2} and Haah's cubic code~\cite{haah} as examples of type-I and type-II fracton orders respectively\footnote{In type-I models, fractons are created at the corners of membrane-like operators, while they are created at the corners of fractal-like operators in type-II models. Type-II phases are also distinct from their type-I counterparts by the absence of \textit{any} topologically non-trivial mobile quasi-particles, while type-I models generally host sub-dimensional particles in addition to fractons.}. We discuss the existence of an edge-ES correspondence for these systems towards the end of this section. 

Although our general derivation of the EH in the previous section did not rely upon the simplifying assumption of considering only CSS codes, all models studied in this section are CSS stabilizer Hamiltonians of the form Eq.~\eqref{eq:CSS}. We also note that while we describe the NLSS below, a more thorough discussion for most cases considered here can be found in Ref.~\cite{albert}. As a matter of notation, if a stabilizer is cut, we refer to it as an $(a,b)$ cut stabilizer, where $a(b)$ is the size of its support in $A(B)$ for a bipartition $(A,B)$ of the qubits forming $\mc{H}$, the Hilbert space of the system in question. Throughout this paper, we will only consider bipartitions associated with the degrees of freedom living in spatially distinct regions $(A,B)$ (with $B = A^c$), such that the boundary between regions $A$ and $B$ defines the \textit{entanglement cut}.

\subsection{$\mathbb{Z}_2$ Topological Order}
\label{toriccode}

\subsubsection{$d=2$ Toric Code/Wen-plaquette Model}

We start with the $d=2$ toric code~\cite{kitaev2} and the Wen-plaquette model~\cite{wenplaquette}, examples of phases hosting $\mathbb{Z}_2$ topological order which are well-known to be unitarily equivalent in the bulk via a local unitary transformation and a $\pi/4$ rotation of the lattice (see e.g. the discussion in Ref.~\cite{wenbook}). Hence, we only discuss the toric code explicitly here, with results for the Wen-plaquette model following immediately. 
The toric code in $d=2$ spatial dimensions is defined on the square lattice with one qubit associated with each edge and is described by the Hamiltonian
\beq
\label{eq:HTC}
H_{TC_{d=2}} = - \sum_v A_v - \sum_p B_p.
\eeq
The first term in the Hamiltonian is associated to every vertex $v$ of the lattice such that $A_v =\prod_{e @ v} X_e$, \textit{i.e.,} the $X$-type Pauli for each edge attached to $v$. The second term is associated with every plaquette $p$ such that $B_p =\prod_{e \in p} Z_e$, \textit{i.e.,} every $Z$-type Pauli forming $p$.

For generic entanglement cuts, any boundary\footnote{To avoid verbosity, and since it should be clear from context, we will refer to a ``boundary defined by the entanglement cut'' as simply a ``boundary,'' except when contrasting it with a physical boundary of the system.} parallel with the coordinate directions contributes to the EH at first order, since both vertex and plaquette stabilizers are either $(3,1)$ or $(1,3)$ cut stabilizers. Analogously, any corners of the entanglement cut contribute at second order as both stabilizer types are $(2,2)$ cut stabilizers; this is also the case for straight boundaries which cut diagonally with respect to the coordinate directions (which is the natural cut for the Wen-plaquette model).

\begin{figure}[t]
    \centering
    \includegraphics[scale=.4]{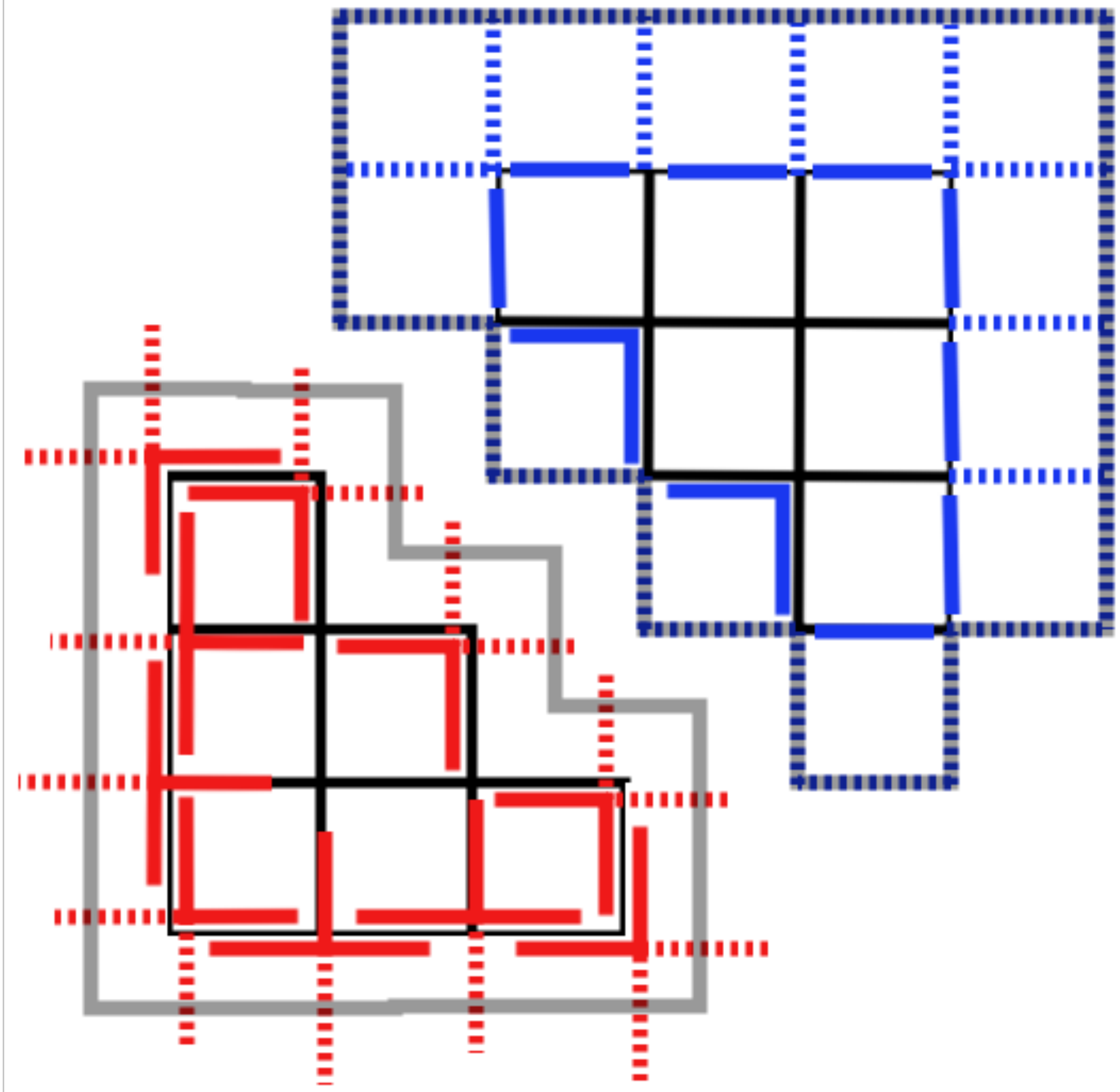}
    \caption{Depiction of the cut surface stabilizers which form the entanglement Hamiltonian for an arbitrary cut of the $d=2$ toric code as well as the NLSS they form. Solid lines represent terms in the EH, while dotted lines represent the completion in $B$ of the cut stabilizers.  Red lines represent $X$-type operator, blue lines represent $Z$-type operators, and the grey lines emphasise the loop operators which form NLSS as represented in $B$.}
    \label{fig:tcbdry}
\end{figure}

%We now consider two boundaries, both of which are depicted in Fig.~\ref{fig:tcbdry}. The first case is along the flat edge which yields the entanglement Hamiltonian
We now consider these two types of local boundaries in detail, both of which are depicted in Fig.~\ref{fig:tcbdry}. The first is the case of a locally flat edge which contributes terms to the EH of the form,
\beq
\hlf \sum_{e_A \parallel \partial} \xi_{e_A}^Z \tilde Z_{e_A} + \hlf \sum_{e_B \perp \partial}\xi_{e_B}^X \tilde T_{v(e_B)} \in H_{\text{ent}}^{TC_2},
\eeq
where $e_A \in A$ ranges over all edges parallel to the boundary $\partial$ of the entanglement cut and $e_B \in B$ ranges over all perpendicular edges. $v(e_B)$ corresponds to the vertex in $A$ which is attached to the edge $e_B$ and $\tilde{T}_v= \left(A_v\right)_A$ is the T-shaped cut vertex stabilizer at $v \in \partial$ (see Fig.~\ref{fig:tcbdry}). The second case is along a ``stair-stepping'' edge; since there are no contributions at first order, to second order we find that the EH contains terms of the form
\beq
 \hlf \sum_{\text{concave }v \in \partial} \xi_v^X \tilde E^X_v + \hlf \sum_{\text{convex }v \in \partial} \xi_v^Z \tilde E^Z_{p(v)} \in H_{\text{ent}}^{TC_2},
\eeq
where $v$ ranges over all vertices on the boundary, and $p(v)$ is the plaquette such its intersection with the boundary is $v$. Here, $\tilde{E}_v^X = \left(A_v\right)_A$ and $\tilde{E}_p^Z= \left(B_p\right)_A$ are the ``elbow''--shaped operators formed by both the vertex and plaquette operators (see Fig.~\ref{fig:tcbdry}). Considering these two possibilities is sufficient to obtain all contributions to the EH coming from any bounded entanglement cut.

Considering the commutation relation between the various terms, we see there exists a mapping from the resulting EH to the $d=1$ transverse-field Ising model (TFIM). In particular, any $X$-type operator (be that an edge or a concave elbow operator) anti-commutes with the two neighboring $Z$-type operators (be they T-shaped or convex elbow operators) and vice versa. These are precisely the commutation relations between the terms of the TFIM. This implies that we can map the spin-$1/2$ degrees of freedom along the boundary of any region $A$ to effective spin-$1/2$ degrees of freedom along a $d=1$ cycle of the same size as the boundary, whereby the EH is mapped onto a TFIM:
\beq
\label{eq:tc2ent}
H_\text{ent}^{TC_2} \simeq \hlf\sum_{\braket{ij}} \xi_{ij}\widetilde{\mc{Z}_i \mc{Z}_j} + \hlf \sum_i \xi_i \tilde{ \mc{X}}_i, 
\eeq
where $\tilde{\mc Z_{i}}, \tilde{\mc X_i}$ are the Pauli operators acting on the effective spin degrees of freedom, $\xi_i, \xi_{ij}$ are determined by the exact form of the mapping, and the tilde represents that we still have to consider the projection due to the presence of NLSS.

For each stabilizer type, there is an NLSS guaranteed by a topological constraint, {\it i.e.} the product of all stabilizers of that type is the identity. Each NLSS takes the appearance of a closed string wrapping the boundary and is given by the product of all stabilizers of that type inside of $A$. These NLSS constrain the terms of the EH such that
\begin{subequations}\label{eq:NLSScons}
\begin{align}
\prod_{e_B\perp \partial}& \tilde T_{v(e_B)}  \prod_{\text{concave }v} \tilde E^X_v = \mc P^0_A,\\
\prod_{e_A \parallel \partial}& \tilde Z_{e_A} \prod_{\text{convex } v} \tilde E^Z_{p(v)} = \mc P^0_A,
\end{align}
\end{subequations}
with the projectors $\mc{P}^0_A$ defined in Sec.~\ref{sec:pert}. This implies that after the mapping onto the TFIM, the effective spins are constrained by
\begin{subequations}
\begin{align}
\prod_{\braket{ij}}& \widetilde{\mc Z_i \mc Z_j} = I,\label{eq:zzcons}\\
\prod_{i}& \tilde X_i = I. \label{eq:xcons}
\end{align}
\end{subequations}
The above constraints imply that the NLSS ``enforce'' a $\mb Z_2$ invariance, or charge conservation, on the EH. Note that the well-known Kramers-Wannier, or self duality of the $d=1$ TFIM, which maps the transverse field terms $\mc{X}$ onto the Ising terms $\mc{ZZ}$, is exact only up to the global constraint, {\it i.e.} Eq.~\eqref{eq:zzcons} is automatically enforced whereas Eq.~\eqref{eq:xcons} is not. However for the EH~\eqref{eq:tc2ent}, the duality is \textit{exact} since both terms of the EH satisfy the constraints as enforced by the NLSS. There is thus an ambiguity in the mapping of the terms which can be understood as a consequence %both of the bulk $\mb{Z}_2$ topological order and 
of the $e-m$ duality in the bulk.

From the preceding discussion, we see that the EH for the $d=2$ toric code shows universal behaviour, since, for generic cuts and for arbitrary weak local perturbations, it can be mapped onto a $\mb{Z}_2$ invariant TFIM of the form Eq.~\eqref{eq:tc2ent}, which acts on a chain of effective spin-$1/2$ degrees of freedom. Importantly, it is the NLSS which enforce the global $\mb{Z}_2$ invariance of the EH, which can thus be understood as a consequence of the bulk $\mb{Z}_2$ topological order. Similarly, we see from the NLSS that the electromagnetic duality of the bulk state manifests itself as Kramers-Wannier duality in the EH, which suggests that there exists a finite region in parameter space for which the EH can be mapped onto the critical Ising CFT. In particular, for entanglement cuts corresponding to a flat boundary, imposing translation symmetry along the boundary forces all $\xi$ coefficients in the perturbation $V$ to be equal, which maps the EH onto the critical 1d Ising CFT. The condition that all $\xi$ coefficients be equal for the stair-stepping edge can be understood as being enforced by translation symmetry along a flat entanglement boundary of the Wen-plaquette model, in which case the EH again maps onto the 1d Ising model at its critical point. Since this is not expected to hold for generic perturbations, we emphasise that the key universal property of the EH is its mapping onto a $\mb{Z}_2$ invariant Hamiltonian acting on effective spin-$1/2$ degrees of freedom, with the $\mb{Z}_2$ invariance enfored by the bulk topological order through the NLSS. We note that our results for the EH of the toric code/Wen-plaquette model are in agreement with those found in Ref.~\cite{ho1}.

Finally, for the case where $A$ is not bounded, but rather spans one direction of the torus, all results remain the same except for the fact that the NLSS get promoted to logical operators. As a consequence, the right-hand sides of Eqs.~\eqref{eq:NLSScons} are replaced by $(-1)^{\ell_1} (-1)^{\ell_2} \mc P_A^{(\ell_1,\ell_2)}$, where $\ell_1, \ell_2$ are the indices for the logical operators. When $A$ was bounded, the projection was necessarily onto the ``zero charge'' sector of the TFIM, but due to the possible negative sign, the projection is now onto the charge sector which corresponds to the topological sector in the bulk. As a consequence, if we are in the 1-charge sector for both topological indices, the $\mc Z \mc Z$ terms must multiply to $-1$, something which is otherwise impossible in the 1d TFIM. To understand this better, consider a flat cut extending along one direction. Even though the product of all $T$ operators has no support on the qubits along the boundary (the hypothetical TFIM degrees of freedom), this operator extends into the bulk and can hence have an eigenstate which differs from one.

\subsubsection{$d\geq 3$ Toric Code}

The most natural, albeit by no means the only, generalisation of the toric code for $d \geq 3$ is defined on the $d$-dimensional Euclidean lattice, with one qubit per edge. The Hamiltonian $H_{TC_d}$ is identical to that of the $d=2$ toric code~\eqref{eq:HTC}, with the stabilizers of the exact same form, where the vertex stabilizer $A_v$ is formed by the $2d$ X-type Pauli operators attached to a vertex and the plaquette stabilizer $B_p$ is still the product of the four Z-type operators forming a plaquette.

\begin{figure}[t]
    \centering
    \includegraphics[scale=.4]{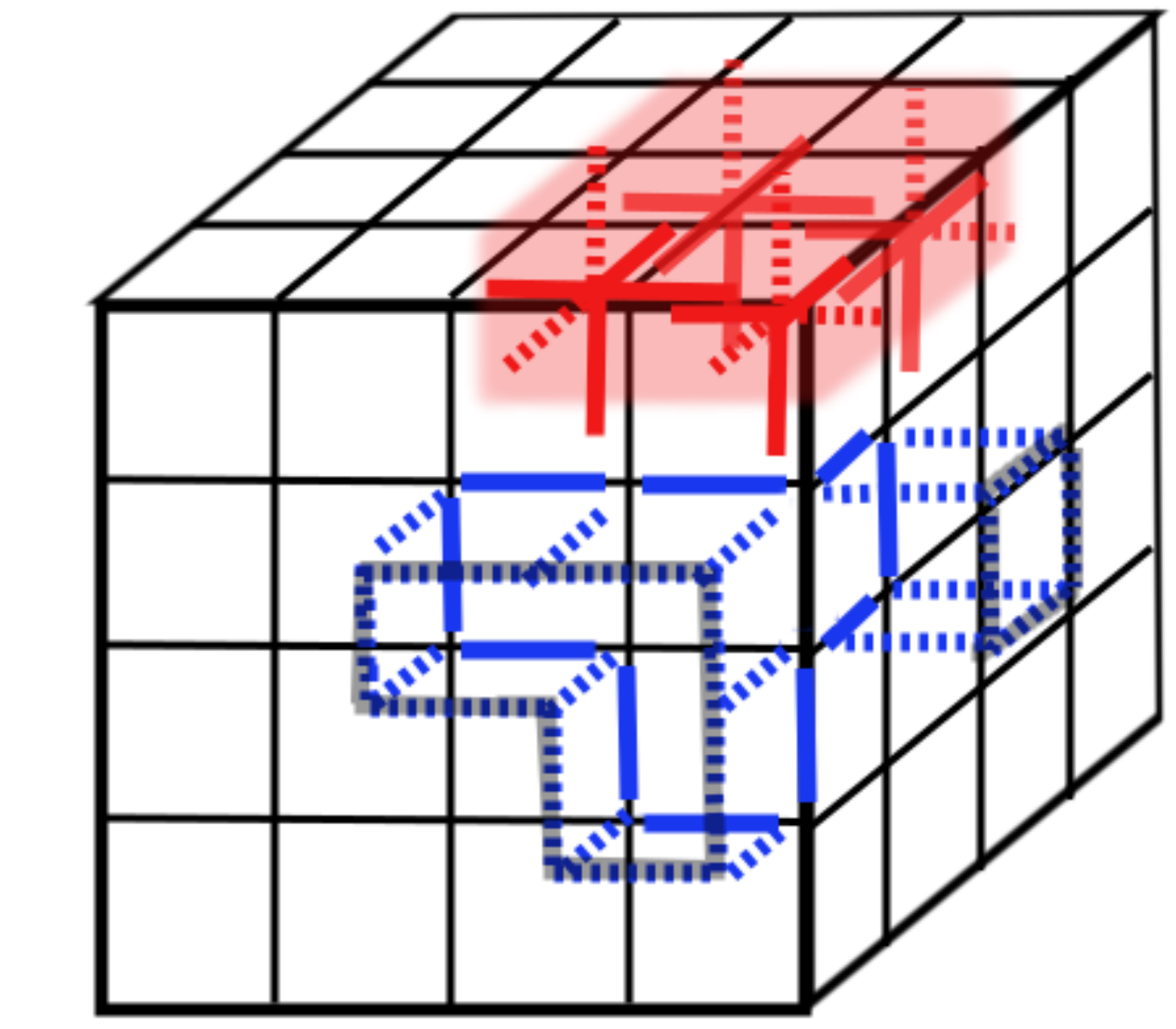}
    \caption{Depiction of the cut surface stabilizers which form the entanglement Hamiltonian for a $4 \times 4 \times 4$ cubic cut of the $d=3$ toric code as well as the NLSS they form. Solid lines represent terms in the EH, while dotted lines represent the completion in $B$ of the cut stabilizers.  Red lines represent $X$-type operator, blue lines represent $Z$-type operators, and the grey lines emphasise the loop operators which form $Z$-type NLSS as represented in $B$. Note the presence of this NLSS is insensitive to the geometry of the hinge. The shaded area emphasises a portion of the wrapping membrane operator which forms the $X$-type NLSS as represented in $B$. Again, the presence of this NLSS is insensitive to the geometry of the hinge. }
    \label{fig:tc3}
\end{figure}

We focus here on the $d=3$ toric code and begin by looking at contributions to the EH stemming from the vicinity of a flat boundary with respect to the coordinate directions. These parts of the boundary have contributions coming from the $(1,3)$ cut plaquettes which intersect the boundary along an edge as well as from the $(5,1)$ cut vertex stabilizers, as depicted in Fig.~\ref{fig:tc3}. These contributions take the form
\beq
 \hlf \sum_{e_A \parallel \partial}\xi_e^X \tilde Z_e + \frac{1}{4} \sum_{e_B \perp \partial}\xi_{e_B}^Z \tilde T_{v(e_B)} \in H_{\text{ent}}^{TC_3}.
\eeq
Along any ``hinges'' of the entanglement cut, we have additional $(2,2)$ cut plaquettes for concave hinges and $(4,2)$ cut vertex contributions at convex hinges, as shown in Fig.~\ref{fig:tc3}. Likewise, any convex corner contributes one $(3,3)$ cut vertex stabilizer whereas concave corners contribute no additional terms. The presence of such hinges and corners in the entanglement cut add terms to the EH of the form:
\beq
\frac{1}{6} \sum_{\text{convex }v \in\partial} \xi_v^X \tilde T_v
 +\frac{1}{2} \sum_{\text{convex } v \in \partial} \xi^Z_{v} E^Z_{p(v)} \in H_{\text{ent}}^{TC_3}.
 \eeq
The terms considered above are sufficient for generating all contributions to the $d=3$ toric code EH coming from any bounded entanglement cut. An example of such contributions are shown in Fig.~\ref{fig:tc3} for a cubic entanglement cut.

To understand the universal features of the EH of the $d=3$ toric code, we now consider the NLSS. In general, there exist an extensive number of NLSS since the local cube constraints ($\prod_{p \in \text{cube}} B_p = I$) imply that every product of $\tilde Z_e$ and $\tilde E^Z_{p(v)}$ which forms a closed loop along the boundary of the cut must equal the identity (see Fig.~\ref{fig:tc3}). Note that this includes closed loops which are deformed around hinges and corners. The same constraint is present among the the $\mc Z \mc Z$ terms of a $d=2$ TFIM. Similarly to the $d=2$ toric code, this suggests a mapping from the original qubits along the boundary onto effective spin-$1/2$ degrees of freedom on the convex vertices of the boundary
\footnote{As a consequence of fixing some degrees of freedom along the boundary---specifically, those associated with the complete plaquette stabilizers---the total Hilbert space dimension of the effective degrees of freedom on the boundary is reduced. This same dimensional reduction is also present in the X-cube model, which we discuss below.}. Correspondingly, at the operator level the mapping is implemented through
\begin{subequations}
\begin{align}
\tilde Z_e, E^Z_{p} \to& \widetilde{\mc Z_i \mc Z}_j, \\
\tilde{T}_v \to& \tilde{\mc X_i},
\end{align}
\end{subequations}
where $e= (i,j)$ for $Z_e$ terms or $i$ and $j$ corresponds to the two adjacent corners of $p$ for $E^Z_p$ terms, and $v=i$. As before, $\tilde{\mc Z}_{i}, \tilde{\mc X}_i$ are the Pauli operators acting on the effective spin degrees of freedom and the tilde represents projection due to the presence of NLSS.

The above mapping preserves all the commutation relations between the different terms of the EH and thus, we find that the EH for the $d=3$ toric code is unitarily equivalent to the $d=2$ TFIM: 
\beq
\label{eq:tc3ent}
H_\text{ent}^{TC_3} \simeq \hlf\sum_{\braket{ij}} \xi_{ij}\widetilde{\mc{Z}_i \mc{Z}_j} + \frac{1}{4} \sum_i \xi_i \tilde{ \mc{X}}_i,
\eeq
where $\braket{i,j}$ now refers to nearest neighbors on the square lattice and $\xi_i, \xi_{ij}$ are determined by the precise form of the mapping. Note that unlike the $d=2$ toric code, we were able to make the mapping explicit in this case as there is no ambiguity in how the different terms get mapped. This is due to the lack of an exact $e-m$ duality in the $d=3$ toric code, which translates into the absence of an exact self-duality in the EH, as evinced also from its equivalence to the $d=2$ TFIM. While the $d=2$ TFIM does not harbour a Kramers-Wannier duality, it is dual to the $d=2$ Ising gauge theory via the celebrated Wegner duality~\cite{wegner}, indicating that the EH for the $d=3$ toric code is dual to a theory with a \textit{local} $\mb{Z}_2$ symmetry. Further, although the existence of a phase transition in the $d=2$ TFIM shows that there exists a region in parameter space for which the EH is at the critical point, generically we do not expect the EH to be near criticality.

Aside from the ``trivial'' NLSS, there are also those connected to the topological constraints in the bulk. For topologically trivial entanglement cuts there is only one such constraint, and thus only one NLSS constraint( $\prod_v A_v =I$) not already captured by the trivial NLSS, namely:
\beq
\prod_{e_B\perp \partial} \tilde T_{v(e_B)}  \prod_{\text{concave }v} \tilde T_v = \mc P^0_A,
\eeq
which for the effective TFIM implies
\beq
\prod_{i} \tilde{\mc X_i} = I.
\eeq
This constraint enforces global $\mb{Z}_2$ invariance on the EH and can be understood as a consequence of charge conservation for the electric sector in the bulk. Much like the EH for the $d=2$ toric code, we hence find that the universal features of the EH for the $d=3$ toric code are its mapping onto a globally $\mb{Z}_2$ symmetric Hamiltonian acting on effective spin-$1/2$ degrees of freedom, with the symmetry enforced by the bulk topological order through the topologically non-trivial NLSS. Where the EH for the $d=2$ toric code harbours a self-duality, for the $d=3$ toric code the EH is dual to the $d=2$ Ising gauge theory. 

So far, it is not clear whether the EH can additionally encode the flux conservation condition in the bulk. In the absence of perturbations, there exist contributions to the entanglement, which appear in the recoverable information, originating from the flux conservation condition. However, this only occurs for topologically non-trivial cuts such that the bulk can enclose or encircle closed flux lines. Since the recoverable information counts non-trivial NLSS, we expect the EH to encode this information as well for topologically non-trivial cuts. We return to this point in Sec.~\ref{sec:summary}. 

We now consider the case when $A$ is not bounded, \textit{i.e.,} when the boundary wraps around the system. For concreteness, we assume the boundary is a flat surface which is everywhere perpendicular to the $z$-direction. Other possibilities can be analogously understood. All of our discussion for bounded cuts carries over with the key difference being the promotion of non-trivial NLSS to logical operators. The electric charge sector gets projected onto the topological sector given by the quantum number of the membrane logical operator perpendicular to the $z$ direction. Likewise, we also find two flux sectors defined by
\begin{subequations}
\begin{align}
\prod_{e\in \text{path}_x} & \tilde Z_e =(-1)^{f_x} \mc P_A^{(f_x, f_y)}\\
\prod_{e\in \text{path}_y} & \tilde Z_e =(-1)^{f_y} \mc P_A^{(f_x,f_y)},
\end{align}
\end{subequations}
where $\text{path}_{x(y)}$ is any path which wraps around the $x(y)$ direction, and $f_{x(y)}$ is the quantum number for the logical string operator wrapping around the $x(y)$ direction. Similarly to our discussion of the $d=2$ toric code, we find that these conditions are only possible since the mapping of the EH onto the $d=2$ TFIM is exact only up to global constraints. 

\subsection{Type-I Fracton Order: X-cube model}
\label{xcube}

\begin{figure}
    \centering
    \includegraphics[width=0.4\textwidth]{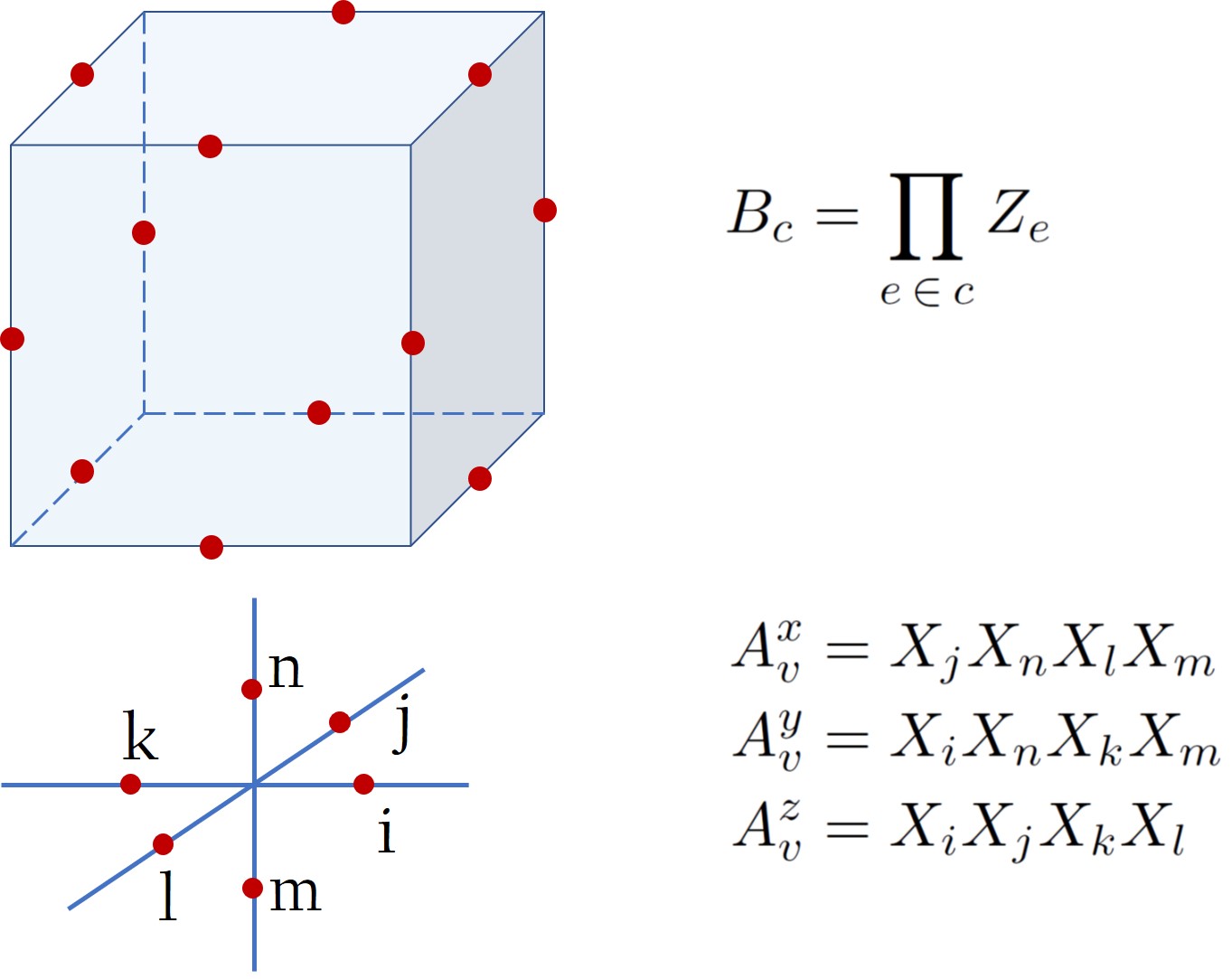}
    \caption{The X-cube Hamiltonian is defined on the cubic lattice with qubits (red dots) placed on each edge and is given by the sum of two stabilizer terms---the first $B_c$, is the product of twelve $Z$-operators at each cube $c$, while the second $A_v^i$ is the product of four $X$-operators attached to each vertex $v$ in the plane perpendicular to the coordinate direction $i$.}
    \label{fig:xcham}
\end{figure}

As the archetypal model displaying type-I fracton order, we now study the X-cube model introduced in Ref.~\cite{fracton2}. The model is defined on a cubic lattice, with qubits living on each edge of the lattice. It is a CSS Hamiltonian formed by two different stabilizer types,
\beq
H_{XC} = -\sum_c B_c -\sum_{v,i} A_v^i,
\eeq
where $i = x,y,z$. The first stabilizer type $B_c$ is associated with each cube $c$ such that $B_c = \prod_{e \in c} Z_e$ \textit{i.e.,} every Z-type Pauli forming the cube. The second stabilizer type has three stabilizers associated to each vertex, one for each direction, such that $A_v^i =\prod_{e@v, \perp i}X_e$ \textit{i.e.,} every X-type Pauli attached to $v$ and in the plane perpendicular to the direction $i$, as depicted in Fig.~\ref{fig:xcham}. Given the extensive literature on fracton order, we do not discuss the properties of this model in detail here but refer the reader to Ref.~\cite{fractonreview} for a review.

Unlike topologically ordered systems, those with fracton order have a non-trivial geometric dependence~\cite{slagle3,shirleygeneral,cagenet,symmetric,gromov2}; for the X-cube model, this is evidenced clearly from its sub-extensive ground state degeneracy on the 3-torus~\cite{fracton2}. Due to the geometric sensitivity, tabulating all possible contributions to the EH from the set of all bounded cuts is much more challenging than for the $d=2,3$ toric codes discussed in the previous section. However, such a tabulation is unnecessary since the relevant features distinguishing the entanglement structure of fracton order from that of topological order can be understood by studying an $R\times R \times R$ cubic cut along the coordinate directions, which we now consider. 

\begin{figure}[t]
    \centering
    \includegraphics[scale=0.4]{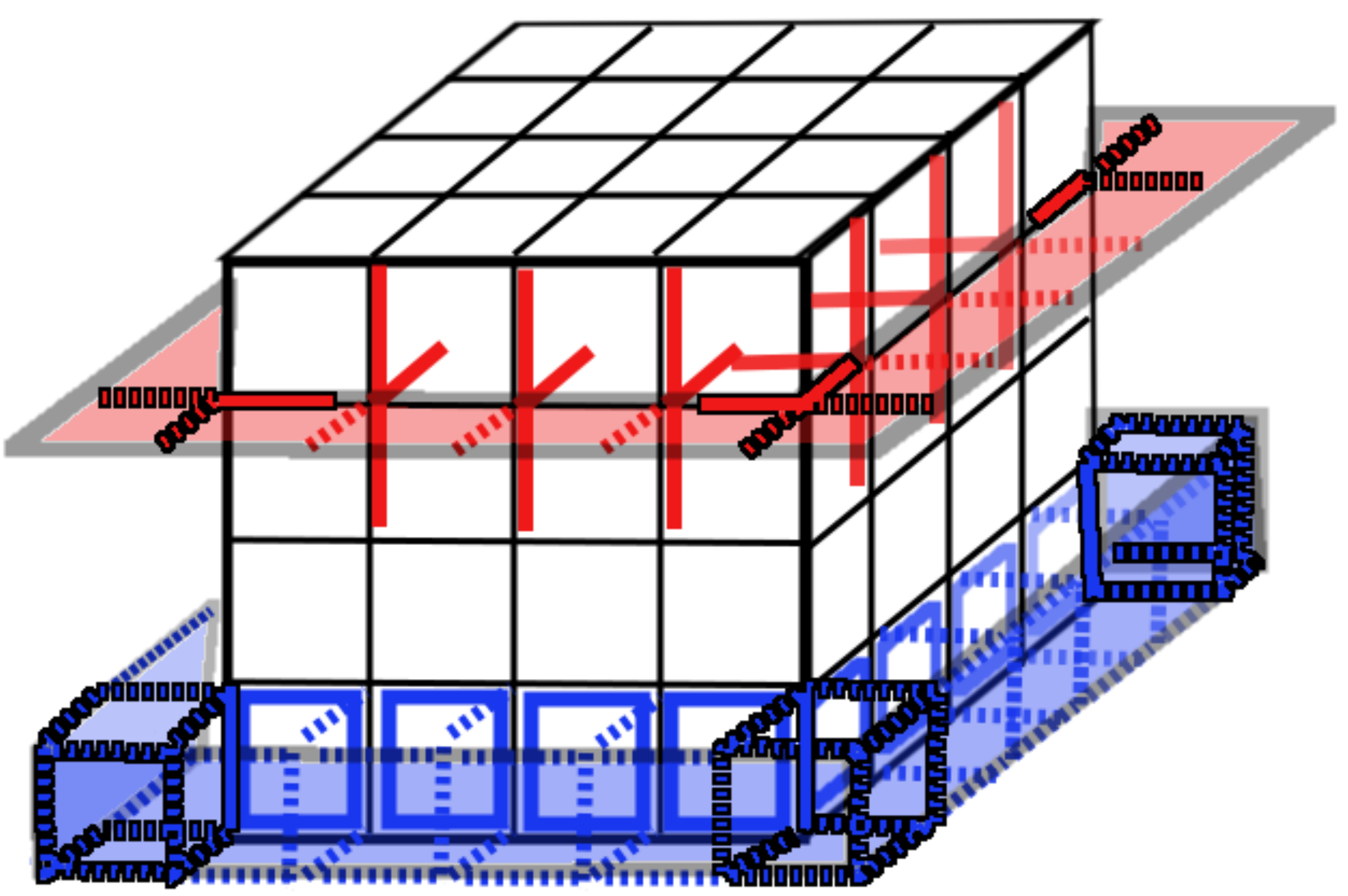}
    \caption{Depiction of the cut surface stabilizers which form the entanglement Hamiltonian for a $4 \times 4 \times 4$ cubic cut of the X-cube model as well as the NLSS they form. Solid lines represent terms in the EH, while dotted lines represent the completion in $B$ of the cut stabilizers.  Red lines represent $X$-type operator, blue lines represent $Z$-type operators, and the shaded areas emphasise the ribbon operators which form NLSS as represented in $B$. Hinge contributions have been highlighted.}
    \label{fig:xcbdry}
\end{figure}

Within the interior of each plane of this cut, but away from its boundaries (hinges and corners), only those $(3,1)$ vertex stabilizers whose directional index is along that plane survive at first order {\it i.e.}, only the $A_v^i$ and $A_v^j$ $(3,1)$ stabilizers survive on the two $ij$ planes of the cut. We must go to fourth-order before we find a $(4,8)$ cut cube stabilizer inside each plane. Both are demonstrated in Fig.~\ref{fig:xcbdry}. Thus the surface of the cut, away from the hinges, contributes
\beq \label{eq:XCentH1}
\frac{1}{4} \sum_{e_B \in \partial} \xi_{e_B}^X \tilde T_{v(e_B)} + \frac{\epsilon}{8} \sum_{p \in \partial} \xi_p^Z \tilde B_p \in H_{\text{ent}}^{XC},
\eeq
where $\tilde{B}_p$ is the same plaquette operator as in the toric code and $\epsilon = \lambda^3$, since the second term appears at fourth order in perturbation theory. Note that we only include one $\tilde T_v$ for each vertex along the boundary as the product of the three vertex terms $A_v^i$ at any given vertex is a constraint. This implies the presence of a trivial NLSS requiring the equivalence of the two cut stabilizers at $v$.

Next, along the hinges of the cut but away from its corners, we find $(1,11)$ cut cube stabilizers which intersect the hinges. Since the cut vertex stabilizers which contribute are the same as those coming from the surface, these have already been accounted for in Eq.~\eqref{eq:XCentH1}. In addition, there are also 
contributions from $(2,2)$ cut vertex (elbow) operators. However, this is a second-order contribution which, due to the superficial NLSS, is the product of two first-order contributions, and is thus excluded. Finally, there are no cube contributions coming from the corners of the cut but there do exist three $(2,2)$ cut vertex stabilizers whose product is the identity. As all three of these appear at the same (second) order, we include all of them. Therefore, we find additional contributions to the EH coming from the hings and the corners, which take the form:
\begin{align}\label{eq:XCentH2}
 \frac{1}{4} \sum_{e_A \in \text{hinge}}\xi_{e_A}^Z \tilde Z_e + \hlf \sum_{ \text{corner } v,i} \xi_{v}^i \tilde E_v^i \in H_{\text{ent}}^{XC},
\end{align}
Eqs.~\eqref{eq:XCentH1} and \eqref{eq:XCentH2} include all possible contributions to the EH, where the hinge contributions are highlighted in Fig.~\ref{fig:xcbdry}.

\begin{figure}[t]
    \centering
    \includegraphics[scale=.2]{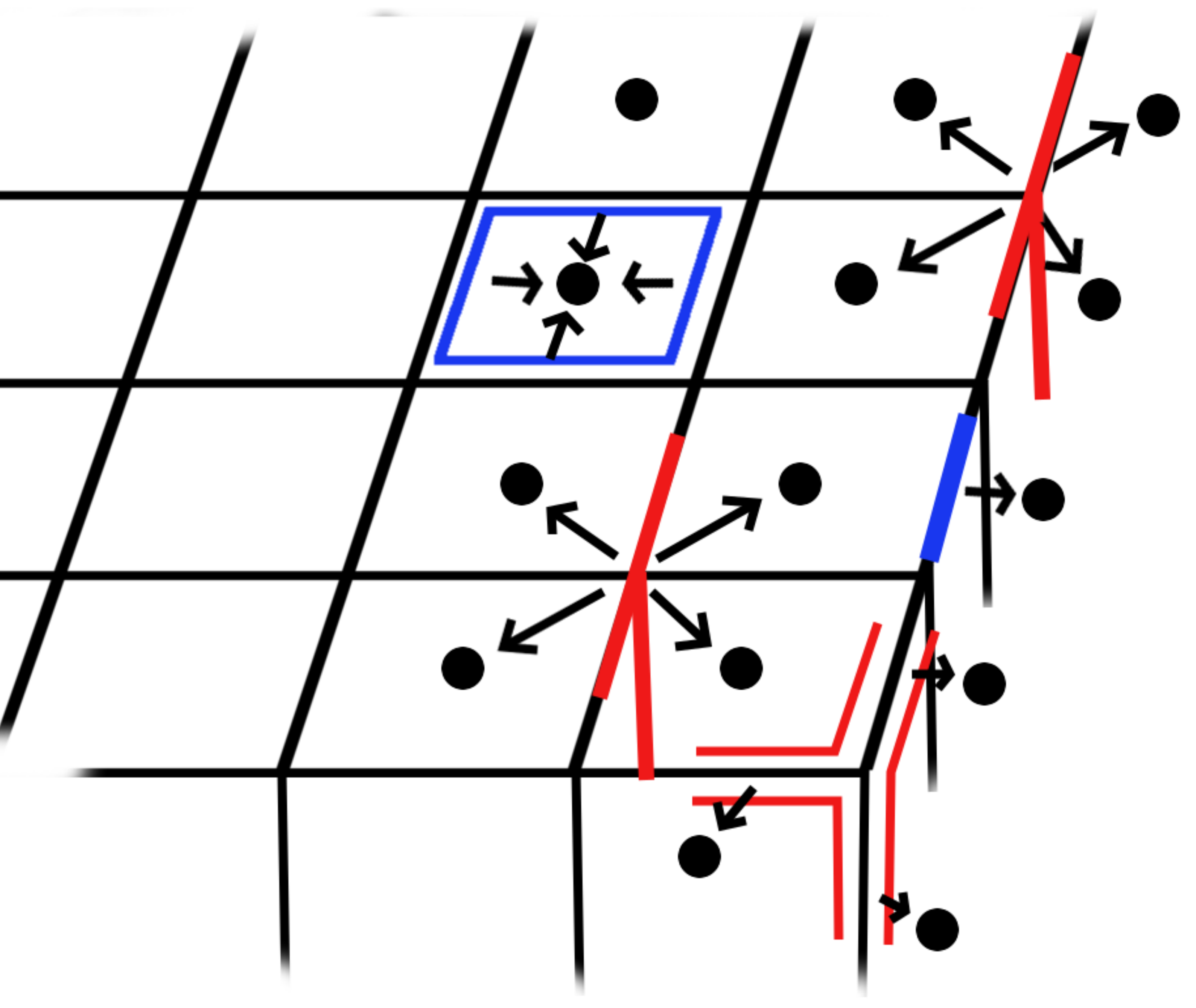}
    \caption{Visual representation of the mapping between the terms of the EH and the effective dual lattice and link degrees of freedom. Red lines represent $X$-type operator and blue lines represent $Z$-type operators on the real degrees of freedom. See text for the definition of operators on the effective degrees of freedom. %\AP{Maybe useful to add dashed black lines showing the dual square lattice on the surface?}
    }
    \label{fig:xcmap}
\end{figure}

Along the surface of the cut but away from hinges and corners, the surface EH, which is formed solely by the terms in Eq.~\eqref{eq:XCentH1}, can be better understood by considering the commutation relations between these terms. Every vertex term $\tilde T_v$ anti-commutes with each of the four plaquette terms $\tilde B_p$ sharing the vertex $v$; likewise, every $\tilde B_p$ term anti-commutes with each of the four vertex terms lying on the corners of $p$. We can then define a mapping which takes the original qubits living on edges of the square lattice on the surface onto effective spin-$1/2$ degrees of freedom living on plaquettes of the original lattice, or equivalently, on vertices of the dual square lattice (see Fig.~\ref{fig:xcmap}). Under this mapping, the $\tilde{T}_v$ and $\tilde{B}_p$ terms of the surface EH~\eqref{eq:XCentH1} are mapped as follows:
 \begin{subequations}
 \begin{align}
 \tilde T_v \to \tilde{\mc B}_q, \\
 \tilde B_p \to \tilde{\mc X}_i,
 \end{align}
 \end{subequations}
where $i$ is the plaquette degree of freedom associated with the $p$, and where
\beq
\label{eq:newplaq}
\tilde{\mc B}_q = \widetilde{\mc Z_{q_1} \mc Z_{q_2} \mc Z_{q_3} \mc Z_{q_4}},
\eeq 
is the product of four (effective) Pauli-$\mc{Z}$ operators acting on the vertices $q_i$ forming the dual plaquette $q$ (see Fig.~\ref{fig:xcmap}). As before, the tilde here signify that we must also account for the constraints imposed by the NLSS.

The surface EH, which on the original square lattice along the surface is given by Eq.~\eqref{eq:XCentH1}, can hence be mapped onto the \textit{subsystem symmetric} transverse field Ising-plaquette (TFIP) model
\begin{align}
\label{eq:XCentH1b}
H_{\text{surface ent}}^{XC} \simeq \frac{1}{4}\sum_q \xi_q \tilde{\mc B}_q + \frac{\epsilon}{8}\sum_i \xi_i \tilde{ \mc X}_i,      
\end{align}
acting on effective spin-$1/2$ degrees of freedom associated with the dual square lattice on the surface of the cut. On the dual lattice, the $\tilde{\mc{B}}_q$ term is given by the product of four Pauli-$\mc{Z}$ acting on the four vertices forming the plaquette $q$. Unlike the $d=2$ TFIM which has a \textit{global} $\mb{Z}_2$ symmetry, the TFIP model instead has a sub-extensive number of $d=1$ \textit{subsystem} symmetries since it is invariant under flipping all spins along any row or column of the lattice. Thus, the EH for the X-cube model is clearly distinct from that of the $d=3$ toric code. 

So far, we have only considered surface contributions to the X-cube EH, but we must also account for the additional hinge and corner contributions given by Eq.~\eqref{eq:XCentH2}. While the EH can be mapped onto TFIP models along the six faces of the cut, these models are coupled through additional effective link degrees of freedom living on the ``wire-frame'' defined by the twelve hinges of the cut. These effective spin-$1/2$ degrees of freedom are then coupled by all terms in Eq.~\eqref{eq:XCentH2} and also by the boundary terms from Eq.~\eqref{eq:XCentH1}. Specifically, at the intersection of any two surfaces $s$ and $s'$ of the entanglement cut, we defined a mapping such that:
 \begin{subequations}
 \begin{align}
 (\tilde T_v)_s \to& \left(\tilde{\mc B}_{(q,l)}\right)_s \nonumber \\ &= \left(\widetilde{\mc Z_{q_1} \mc Z_{q_2}}\right)_s \otimes\left(\widetilde{\mc Z_{l_1} \mc Z_{l_2}}\right)_{\text{link}} \otimes I_{s'},\\
 (\tilde T_v)_{s'} \to& \left(\tilde{\mc B}_{(q,l)}\right)_{s'}\nonumber \\ &= I_s\otimes\left(\widetilde{\mc Z_{l_1} \mc Z_{l_2}}\right)_{\text{link}} \otimes  \left(\widetilde{\mc Z_{q_1} \mc Z_{q_2}}\right)_{s'} \\
 \tilde Z_e \to& I_s \otimes \left(\tilde{\mc X_l}\right)_{\text{link}} \otimes I_{s'},
 \end{align}
 \end{subequations}
 where $\tilde{\mc{B}}_q$ was defined in Eq.~\eqref{eq:newplaq}. Finally, for corners of the entanglement cut (where three surfaces intersect), the mapping is specified by
 \beq
 \tilde E_v^{i} \to \left(\widetilde{ \mc Z_{l_j} \mc Z_{l_k}}\right)_{\text{link}},
 \eeq
 where $ i\neq j\neq k$. %\AP{\sout{All relations for the indices are depicted in Fig.~\ref{fig:xcmap}} 
 The above mappings are depicted schematically in Fig.~\ref{fig:xcmap}.
 
 It is straightforward to show that these mappings preserve all of the commutation relations between the terms in the EH defined by Eqs.~\eqref{eq:XCentH1} and~\eqref{eq:XCentH2}. Considering only their support on the effective link variables, the hinge terms can be mapped onto the TFIM, with the exception that there are two (three on the corners) distinct $\mc Z \mc Z$-like Ising nearest-neighbour terms and, crucially, that the model is defined on the ``cage'' (wire-frame) topology instead of on a cycle. The link variables along the cage are coupled to the surface degrees of freedom by the fact that the $\mc{Z} \mc{Z}$-like terms along the cage anti-commute with the $ \mc X_i$ surface terms. Recall, however, that the $\mc X_i$ terms in the surface TFIP model Eq.~\eqref{eq:XCentH1b} occur at higher order in perturbation theory than the $\mc B$ terms, as indicated by the coeffecient $\epsilon$ of the transverse field term. Thus, to lowest order, we find that the hinge and corner contributions to the EH are mapped onto the TFIM acting on effective spin-$1/2$ degrees of freedom living on the edges of the cubic cut. In addition to distinguishing it from the $d=3$ toric code, the EH also clearly distinguishes the fracton phase from stacks of weakly coupled $d=2$ toric codes, for which the EH would consist of independent TFIMs along each row and column of the square lattice on every surface of the cubic cut. The distinction between the fracton phase and $d=3$ or decoupled stacks of $d=2$ topological orders is also evident in the ``cage-like'' nature of the EH. In the future, it would be interesting to study the evolution of the entanglement spectrum under the flux-string condensation discussed in Refs.~\cite{han,sagar}, whereby stacks of $d=2$ toric codes are strongly coupled to arrive at the X-cube.
 
Following this analysis, we can generalise to any entanglement cut given by a rectangular prism. The X-cube EH can be mapped onto the TFIP model along the surface of the cut, with the hinge and corner contributions mapping onto the TFIM along the cage formed by the edges of cut. Recall that for topologically ordered phases, we found that the EH is universally given by a $\mb{Z}_2$ invariant Hamiltonian acting on effective spin-$1/2$ variables for generic entanglement cuts. Given the geometric nature of fracton order, it should be no surprise that the EH is correspondingly sensitive to the geometry of the entanglement cut. To wit, the EH of the X-cube model can be mapped onto a TFIP model along the surface of the cut and to the TFIM along the hinges of the cut only for cuts which respect the $d=2$ planar subsystem symmetry of the X-cube. For instance, for a cut along the [111]-direction the EH will not take this simple form, and so generically, the EH will not fall into the class of subsystem symmetric invariant Hamiltonians acting on effective qubits---it is only for cuts respecting the $d=2$ planar subsystem symmetry that the EH will be invariant under $d=1$ subsystem symmetries.
 
Returning to a cubic entanglement cut, there is an ambiguity in the mapping of terms in the EH due to the self (or Kramers-Wannier) duality of the TFIP model, similar to what we observed for the $d=2$ toric code. This duality is related to our choice to place effective degrees of freedom on the dual lattice on the surface whereas we could equally have placed them on the vertices of the original lattice~\footnote{We choose to use the dual lattice description here since this makes the hinge and corner mappings appear more natural.}. As with the $d=2$ toric code, the duality for the X-cube EH becomes exact once we consider the NLSS constraints, whereas for a two-dimensional TFIP model the duality is only exact up to the subsystem NLSS constraints along any rigid line. For the effective TFIP model onto which the EH maps, these constraints are primarily responsible for the fractonic behavior among the $\mc B_p$ terms in the EH~\eqref{eq:XCentH1b}. As discussed in Ref.~\cite{albert}, all $6R+1$ independent topological NLSS for the cubic cut are guaranteed by the existence of a sub-extensive number of topological constraints. The constraints amongst the cubic stabilizers are generated by the product of all cubes which contain vertices in any given plane perpendicular to a coordinate direction. Similarly, the topological constraints among the vertex stabilizers is generated by the product of all vertex stabilizers which are entirely supported in any coordinate plane. Thus the product of all stabilizers in $A$ for any one of these planes forms a ribbon NLSS as depicted in Fig.~\ref{fig:xcbdry}. Note that the cube ribbon parallel to the surface is supported on the edge where the ribbon bends. We can characterise the resulting NLSS constraints as
 \begin{subequations}
 \begin{align}
\prod_{v \in \text{ribbon}_\perp} \tilde T_v =& \, \mc P_A^0, \\
\prod_{p \in \text{ribbon}_\parallel} \tilde B_p \prod_{e \in \text{hinge}\,\cap\,\text{ribbon}_\parallel} \tilde Z_e =& \, \mc P_A^0,
 \end{align}
 \end{subequations}
where $\text{ribbon}_\perp$ is any rigid string of vertices wrapping around the entanglement cut along coordinate directions (thus corresponding to a ribbon perpendicular to the surface) and $\text{ribbon}_\parallel$ is a rigid ribbon of plaquettes and edges wrapping around the cut along coordinate directions. After the mapping onto an effective TFIP model, the constraints among the terms of the EH are given by
 \begin{subequations}
 \begin{align}
\prod_{q \in \text{ribbon}_\perp} \tilde{\mc B}_q \prod_{(q,l) \in \text{hinge} \,\cap\, \text{ribbon}_\perp}\tilde{ \mc B}_{(q,l)}=& I, \\
\prod_{i \in \text{ribbon}_\parallel} \tilde{\mc X}_i \prod_{l \in \text{hinge} \,\cap\, \text{ribbon}_\parallel} \tilde{\mc X}_l=& I. 
 \end{align}
 \end{subequations}
The first (plaquette) constraint is naturally enforced whereas the transverse field constraint is not and must be enforced by hand; since this term occurs at higher order in perturbation theory, however, we need only focus on the first constraint. This constraint is responsible for a $\mb{Z}_2$ {\it subsystem} charge conservation which is a key signature of fractonic behaviour~\cite{albertgauge,yizhi1,twisted}, \textit{i.e.,} for the X-cube model, charge is conserved globally but also along each plane. For the TFIM along the hinges of the cut, notice that the product of all parallel ribbon NLSS along any one direction forms a cage. It is this NLSS that constrains the hinge TFIM such that
\begin{align}
\prod_{e \in \text{hinge}} \tilde Z_e = \mc P_A^0,
\end{align}
so that after the mapping we have
\begin{align}
\prod_l \tilde{\mc X}_l =I.
\end{align}
The $\mb{Z}_2$ charge conservation is thus enforced for the entire cage formed by the hinges of the cut and not just for the boundary of a single surface. Finally, we note that any parallel ribbon NLSS which straddles the outer edge of a surface satisfies the definition of a superficial NLSS as all the cut cube stabilizers in this product are along the surface. This lends credence to our assertion that superficial NLSS signal the presence of subsystem symmetries where the superficial NLSS is formed. However, the existence of superficial NLSS is more pertinent to the discussion of Haah's cubic code and SSPT models, so we postpone a detailed discussion of superficial NLSS until Section~\ref{haahcode} and \ref{sec:sspt}. 

\subsection{Type-II Fracton Order: Haah's Cubic Code}
\label{haahcode}

\begin{figure}
    \centering
    \includegraphics[width=0.4\textwidth]{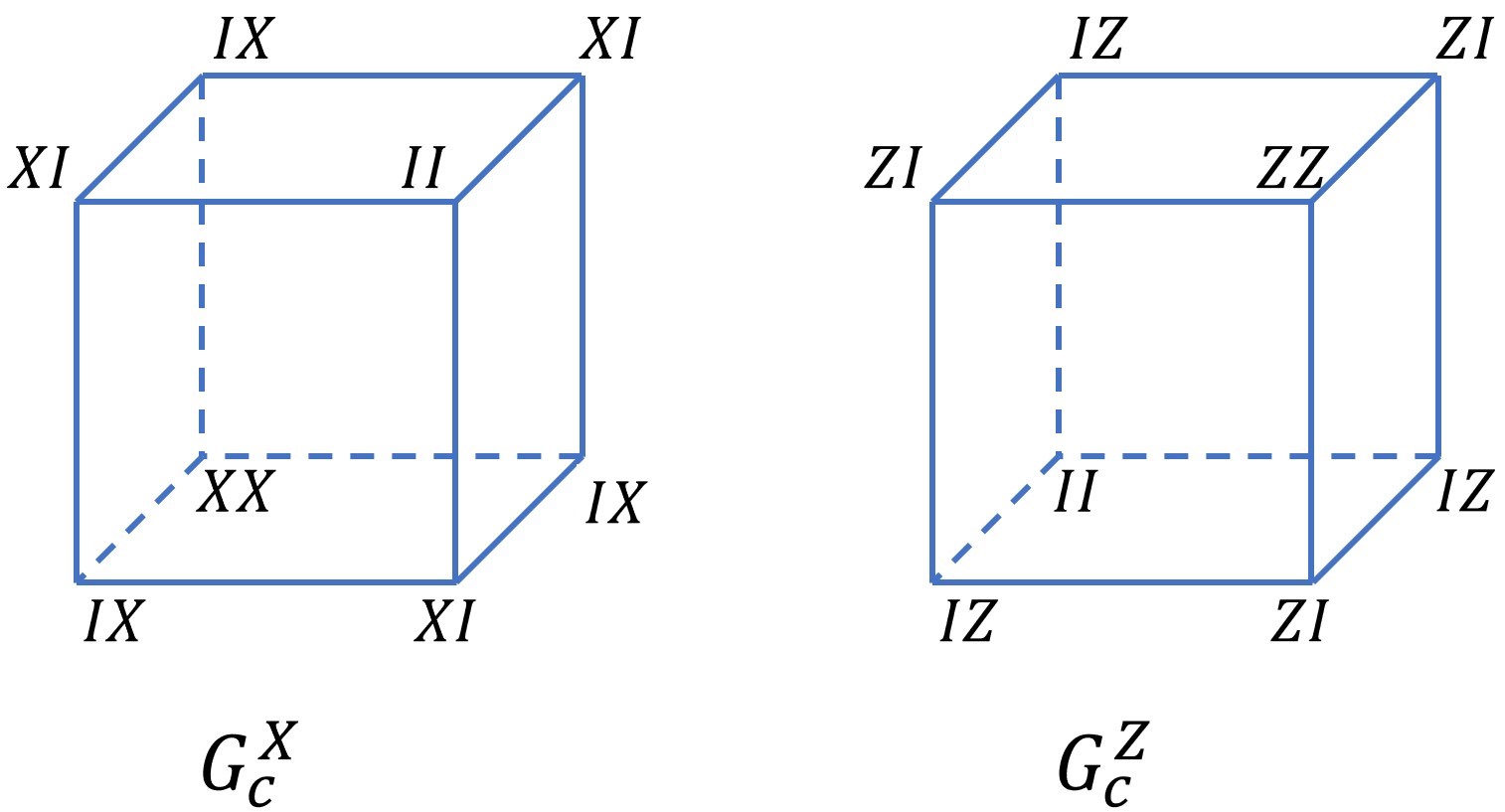}
    \caption{Definition of the stabilizers in Haah's cubic code. The remaining stabilizers are related to these by translations.}
    \label{fig:haahdef}
\end{figure}

We now study the entanglement spectrum of a type-II fracton model, exemplified by Haah's cubic code~\cite{haah}. The model is defined on a cubic lattice with two qubits living on each vertex of the lattice. The model is described by the Hamiltonian
\beq
H_{\text{Haah}} = -\sum_c G_c^X - \sum_c G_c^Z,
\eeq
which consists of two stabilizer types, both of which are associated with every elementary cube of the lattice. The first type is composed of X-type Pauli operators, $G_c^X$, and the other is composed of Z-type Pauli operators $G_c^Z$, with their precise form shown in Fig.~\ref{fig:haahdef}. The curious properties of this model are reviewed in Ref.~\cite{fractonreview}.

\begin{figure}
    \centering
    \includegraphics[scale=0.47]{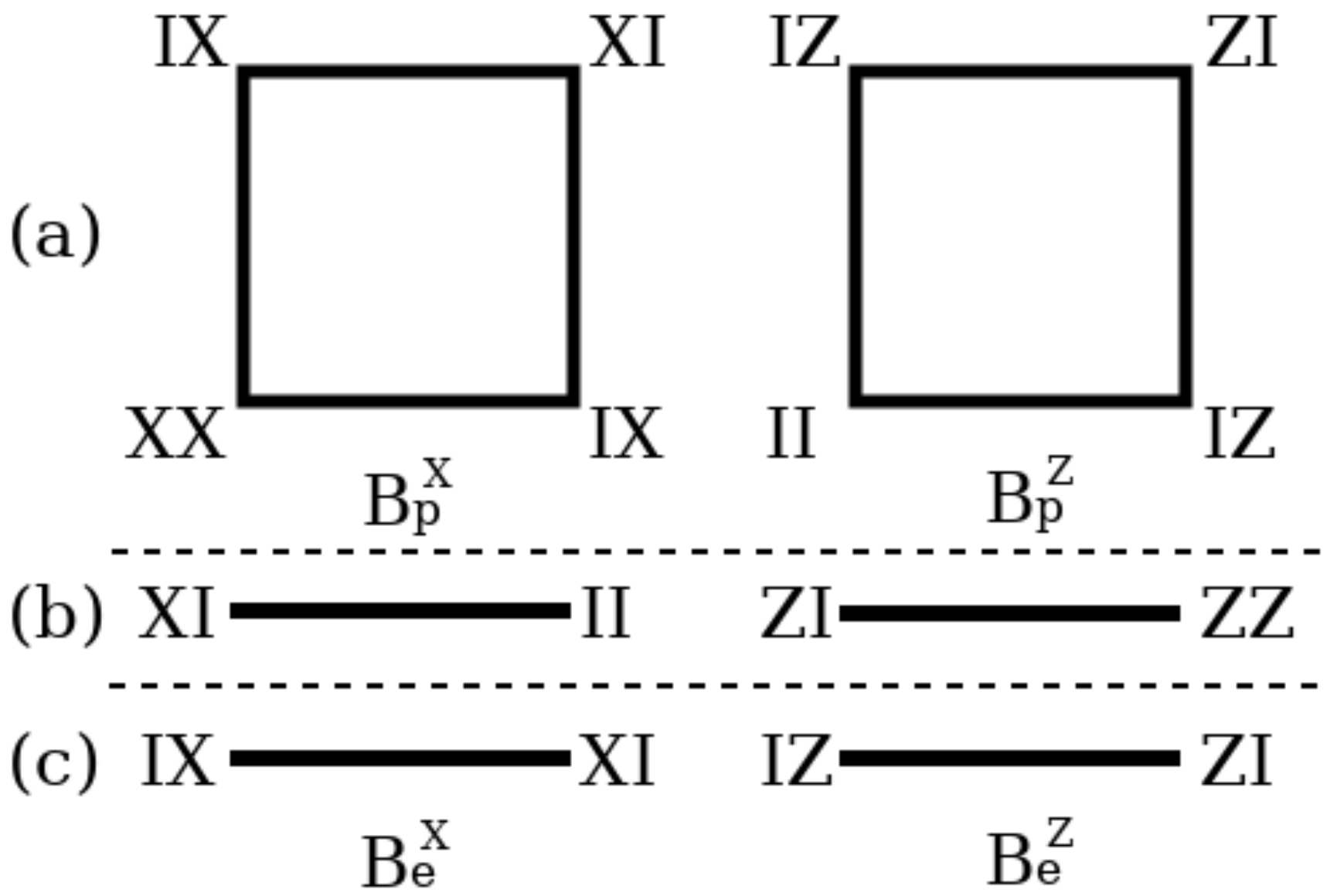}
    \caption{Terms of the entanglement Hamiltonian for the cubic cut for Haah's cubic code. (a): Terms along surfaces of the entanglement cut, away from hinges and corners. (b) and (c): Terms along the hinges of the cut. Each model is defined along six of the twelve edges.}
    \label{fig:haahbdry}
\end{figure}

We start by considering an $R\times R \times R$ cubic entanglement cut along the coordinate directions of the lattice, so as to contrast type-II fracton order with type-I. Given the form of the stabilizers as shown in Fig.~\ref{fig:haahdef}, we require at least third-order perturbations before any contributions arise in the EH along the surface of the cut. These contributions result from the $(3,5)$ or $(5,3)$ cut stabilizers, coming either from $G_c^X$ or $G_c^Z$ depending on which surface we are considering. Regardless of which surface we consider, away from the hinges of the entanglement cut there are two types of terms in the EH for each plaquette of the square lattice on the surface. One such example is depicted in Fig,~\ref{fig:haahbdry}(a), with all others unitarily equivalent. Contributions to the EH from the surface of the cut $s$ are hence give by 
 \begin{align}
 \label{eq:haahsurf}
 \frac{1}{6}\sum_s \sum_{p\in s}\left(\xi_p^X\tilde{B}_p^X + \xi_p^Z \tilde{B}_p^Z\right) \in H_{\text{ ent}}^{\text{cube Haah}},
\end{align}
 where $\tilde{B}_p^X = \left(G_c^x\right)_A$ or $\left(G_c^x\right)_B$ depending on the particular surface under consideration, and likewise for $B_p^Z$.
 
Along the hinges of the entanglement cut, we find contributions at first, second, and third order coming from the $(1,7),(2,6)$ and $(3,5)$ cut stabilizers which intersect along the hinge (see Fig.~\ref{fig:haahbdry}). The hinges contribute terms of the form:
\begin{align}
 \frac{1}{6} \sum_{e\in \text{hinge}}\left(\xi_e^X\tilde{B}_e^X + \xi_e^Z \tilde{B}_e^Z\right) \in H_{\text{ ent}}^{\text{cube Haah}},
\end{align}
where $\tilde{ B}_e^{X(Z)}$ are defined the same way as the surface terms. The precise form of the cut stabilizer depends on whether the $XX$, $ZZ$, or $II$ corner of the $G_c^X, G_c^Z$ stabilizers are included in the cut edge. For a given edge, if an $II$ corner is included along the edge of one stabilizer type, the $XX$ or $ZZ$ corner is included along the edge of the other stabilizer type, as can be easily checked. 

Unlike all cases considered heretofore, there does not appear to be a familiar Ising-like model onto which the surface EH maps. In fact, we could have anticipated that would be the case given the discussion regarding the role of subsystem symmetries in the our analysis of the X-cube model, where the surface EH maps onto the TFIP model \textit{only} for entanglement cuts along the planar subsystem symmetries. Similarly, we expect that the EH for cuts respecting the fractal subsystem symmetry of Haah's code will be mapped onto a subsystem symmetric Ising-like model---this is indeed the case for a [111] cut, as we discuss later in this section.

Along the hinges of a cubic entanglement cut however, we do find mappings of the EH onto familiar models as long as we ignore any commutation with the surface terms. By considering the mutual commutation relations between the hinge terms specified in Fig.~\ref{fig:haahbdry}(b), we find that these contributions to the EH can be mapped onto the $d=1$ TFIM. This is the case for the six hinges which contribute $(1,7)$ and $(3,5)$ cut stabilizer terms. In contrast, contributions from the remaining six hinges coming from $(2,6)$ cut stabilizers can be mapped onto a CSS version of the $d=1$ cluster model \textit{i.e.,} onto the terms given in Fig.~\ref{fig:haahbdry}(c)~\footnote{To see this, coarse-grain the $d=1$ chain such that there are two qubits per unit cell. After this coarse-graining, there remain only two stabilizers types which are no longer equivalent, up to translates. These can then be mapped via a local Clifford circuit consisting of an operator for each unit cell, such that the resulting stabilizer code corresponds to Fig.~\ref{fig:haahbdry}.}. Finally for the cubic cut, we note in passing that along both the surface and the hinges, there exists a self-duality whereby all $X$-type terms can be exchanged with the $Z$-type terms, leaving the EH invariant.

Although all NLSS are not well-understood for Haah's code, for the cubic cut there exist $12R - 2$ independent NLSS~\cite{albert}. Fourteen of these can be generated using the topological constraints found in Ref. \cite{albertgauge}~\footnote{The ``star'' pattern in Fig 5a of Ref.~\cite{albertgauge} has two independent versions given by exchanging the configurations in the three $[111]$ layers of the triangular lattice over which the pattern is periodic; see the reference for more details.} For example, two are implied by the topological constraints given by the product over all stabilizers of a single type, which enforces
\begin{subequations}
\begin{align}
\prod_{p} \tilde{B}_p^X \prod_{e \in \text{hinge}} \tilde{B}_e^X =\mc P_A^0, \\
\prod_{p} \tilde{B}_p^Z \prod_{e \in \text{hinge}} \tilde{B}_e^Z =\mc P_A^0.
\end{align}
\end{subequations}
Besides the remaining independent NLSS implied by the other twelve known constraints, one can also show that there are no superficial NLSS along any single surface of the cubic cut.

\begin{figure}[t]
    \centering
    \includegraphics[width=0.5\textwidth]{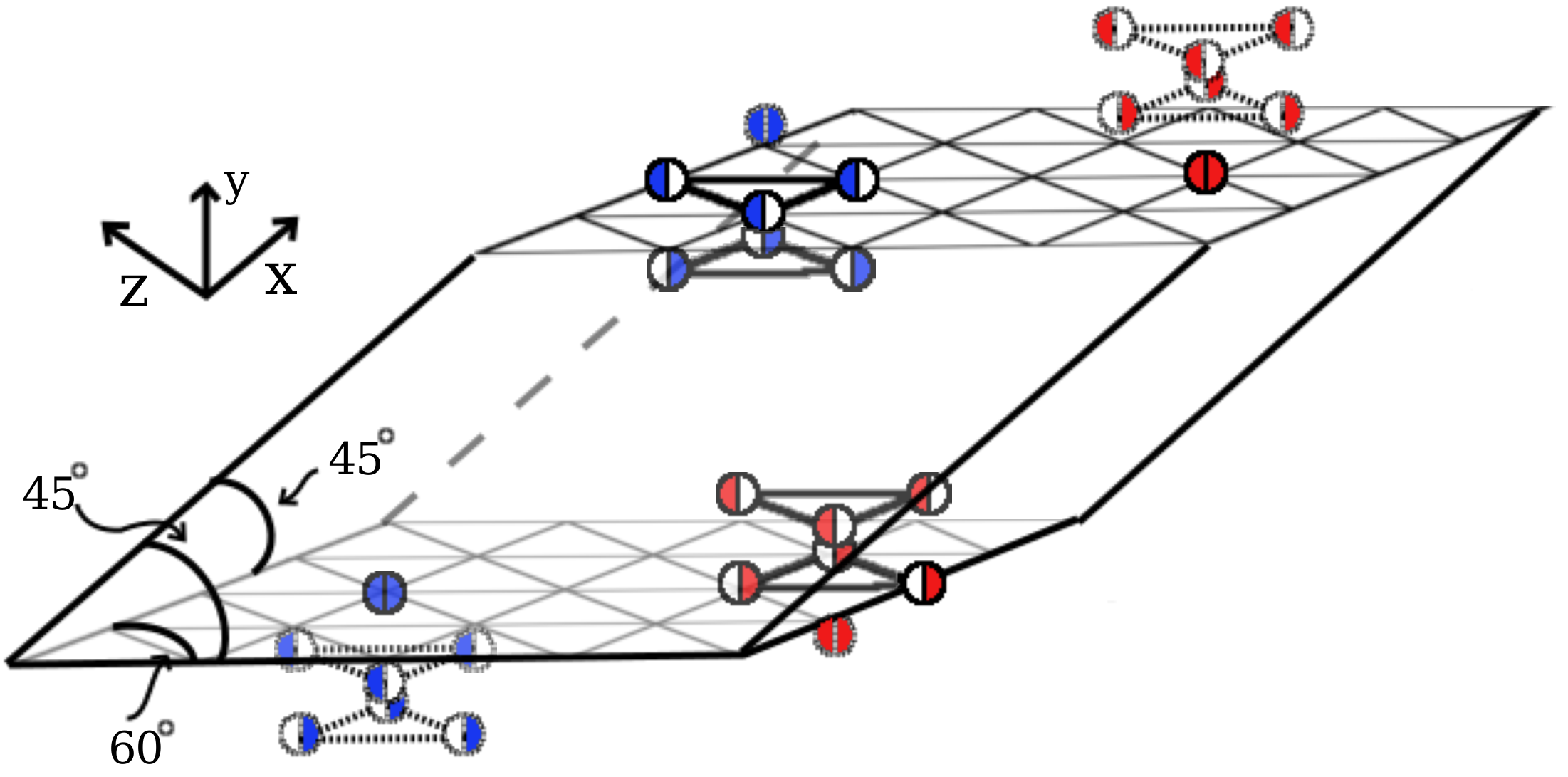}
    \caption{Depiction of the $4 \times 4\times 4$, $45 \degree, 45 \degree, 60 \degree$ parallelepiped cut as with example cut stabilizers along the $[111]$ surface. Solid lines represent terms in the EH, while dotted lines represent the completion in $B$ of the cut stabilizers.  Red semi-circles represent $X$-type operator and blue semi-circles represent $Z$-type operators, where each circle represents the local 2 qubit Hilbert space at a vertex.}
    \label{fig:111cut}
\end{figure}

As anticipated, the cubic entanglement cut does not reveal much about the entanglement structure of Haah's code since it does not respect the (fractal) subsystem symmetries, which played a significant role in our analysis of the X-cube model. Indeed, there exist more illuminating cuts for which we find contributions to the EH at first and second order along the surfaces of the cut. In particular, this is the case for flat entanglement surfaces perpendicular to the $[\pm1,\pm1,\pm1]$ directions, as these cut along the corners of the cubic stabilizers. As a consequence of the three-fold rotational symmetry about the $[111]$ direction and previous results discussed in Ref.~\cite{albertgauge}, the most interesting case is that of the surface perpendicular to the $[111]$ direction. Specifically, let us consider an $R \times R \times R$, $45 \degree, 45 \degree, 60 \degree$ parallelepiped, where the plane of the $60 \degree$ angle is perpendicular to $[111]$ and half of the parallelepiped unit cell forms a corner of a unit cube, as depicted in Fig.~\ref{fig:111cut}. As four of the surfaces for this cut correspond to faces of a cubic cut, the contribution to the EH from those faces is the same as that in Eq.~\eqref{eq:haahsurf}. However, each of the $[111]$ surfaces form a triangular lattice and have contributions distinct from those discussed prior. We look at the surface farthest from the orgin, with results for the closer face obtained by swapping stabilizer types. This surface contains $(2,6)$ $Z$-type and $(6,2)$ $X$-type cut stabilizers, such that its contributions to the EH are given by: 
\beq \label{eq:haahEH}
\frac{1}{6} \sum_{v \in \partial_{[111]}}\xi_v^z\widetilde {\left(Z_v^1 Z_v^2\right)} + \frac{1}{6} \sum_{\triangle \in \partial_{[111]}} \xi_\triangle^x \widetilde {TT}_{\triangle} \in H_{\text{ent}}^{[111]\text{ Haah}},
\eeq
where $\widetilde{TT}_{\triangle} = (G_v^X)_A$ is formed by two stacked triangle operators as shown in Fig.~\ref{fig:111cut} and where $\triangle$ corresponds to the vertex $v$ just above the triangluar plaquette in the $[111]$ direction. Here, the sum is restricted to only one set of triangles, \textit{i.e.,} it only goes over upward-pointing triangles.

Within the boundary layer of $A$, the $TT_\triangle$ operator forms a single triangle operator, strongly reminiscent of the Newman-Moore model~\cite{newmanmoore}. This model is defined on a $d=2$ triangular lattice, with the Hamiltonian given by
\beq
\label{eq:HNM}
H_{NM} = \sum_{i\,,\,j\,,k\,\in\,\triangle} Z_i Z_j Z_k, 
\eeq
where the sum runs over all sets of nearest-neighbour spins $i,j,k$ living on the three vertices of one of the upward-pointing triangles. To make the connection of the $[111]$ surface EH to the Newman-moore model precise, we introduce one effective spin-$1/2$ degree of freedom for every two-qubit unit cell along the entanglement surface and map the terms as follows:
\begin{subequations}
\begin{align}
\widetilde{TT}_{\triangle} \to \tilde{\mc T}_{\triangle}, \\
\widetilde {\left(Z_v^1 Z_v^2\right)} \to \tilde{\mc X_i} ,
\end{align}
\end{subequations}
where $\tilde{\mc T}_{\triangle} =\widetilde{{}\mc Z_{\triangle_1}\mc Z_{\triangle_2}\mc Z_{\triangle_3}}$ such that $\triangle_i$ is one of the effective spin-$1/2$ variables forming $\triangle$ and where $v=i$. The $[111]$ surface Hamiltonian is then mapped onto 
\begin{align}\label{eq:111haah}
H_{\text{ent}}^{[111]\text{ Haah}} \simeq \frac{1}{6} \sum_{\triangle} \xi_\triangle \tilde{\mc T}_{\triangle} + \frac{1}{6} \sum_i \xi_i \tilde{\mc X_i},
\end{align}
which we recognise as the Newman-Moore model Eq.~\eqref{eq:HNM} in the presence of a transverse field. Hence, in contrast with the cubic entanglement cut, for the cut depicted in Fig.~\ref{fig:111cut} we find a natural mapping of the EH for Haah's code onto a transverse field Newman-Moore model, acting on effective spin-$1/2$ degrees of freedom. Taken alongside our results for the X-cube model, this clearly illustrates the non-trivial geometric dependence of fracton phases, whose EH is not universally equivalent to some effective spin model, which is the case for topologically ordered phases. Instead, for both type-I and type-II fracton phases, it is only for specific entanglement cuts \textit{i.e.,} those compatible with the subsystem symmetries of the phase, that the EH maps onto a subsystem symmetric model. For Haah's code, the mapping of the $[111]$ surface EH is onto the Newman-Moore model Eq.~\eqref{eq:111haah}, which is invariant under fractal subsystem symmetries~\cite{yoshida,williamson,devakulfractal}. Our results hence illuminate the crucial role played by subsystem symmetries in the entanglement structure of fracton phases.

\begin{figure}[t]
    \centering
    \includegraphics[scale=.35]{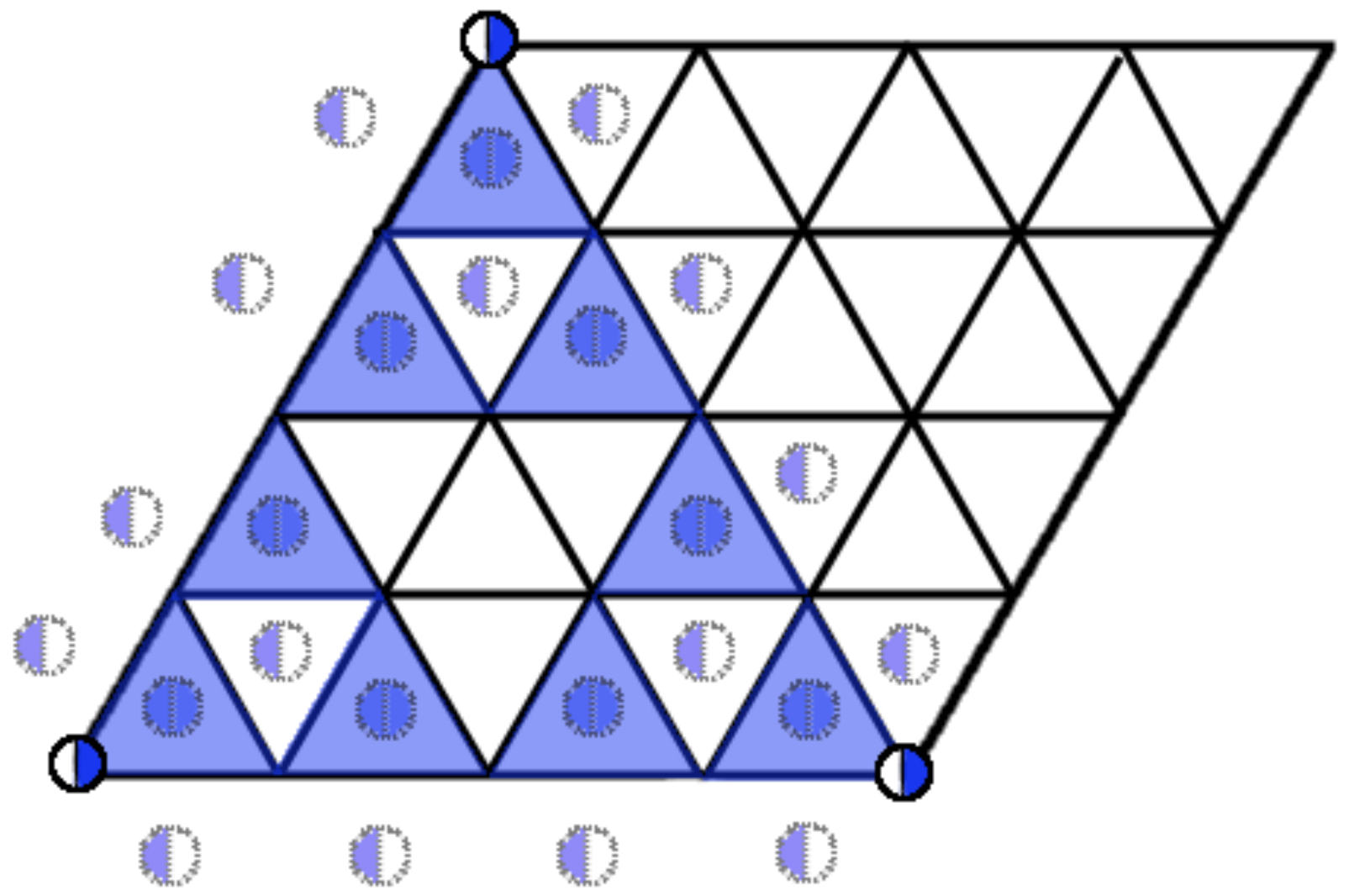}
    \caption{Visual representation of the fractal superfical NLSS for Haah's code along the $[111]$ surface. Solid lines represent the support of the operator in $A$, while dotted lines represent the completion of the operator in $B$. All such operators lie in the $[111]$ layers above the one depicted. Blue semi-circles represent $Z$-type operators, where each circle represents the local 2 qubit Hilbert space at a vertex. The blue shaded triangles represent the stabilizers which form the NLSS. Note if the lattice shown is in $A$ then this is not a full NLSS as there is support in $A$. However, the NLSS is formed by including the three cut stabilizers at the corners of the fractal. These are not included to aid in visualisation.}
    \label{fig:fractalNLSS}
\end{figure}

As before, we still need to consider the effect of the projection due to the NLSS. In Appendix~\ref{ap:RI}, we find that the number of independent NLSS for the cut considered in Fig.~\ref{fig:111cut} are $12R-2$. Fourteen of these are related to the same topological constraints as discussed for the cubic cut and can be explicitly derived from them~\cite{albertgauge}. Unlike a cubic entanglement cut however, for which there are no superficial NLSS along the surfaces, in this case there do exist superficial NLSS along the $[111]$ surface. Consider the $(3,5)$ cut $Z$-type stabilizer such that its support forms a triangle in the boundary layer of $A$. As with the Newman-Moore model, the product of triangle stabilizers forming a Sierpinski fractal only has support on the triangles forming the corners of the fractal. For a fractal of size of the $[111]$ surface, all support is removed from $A$, leaving such a stabilizer group member in $G_B$, as shown Fig.~\ref{fig:fractalNLSS}. This does not affect the terms in Eq.~\eqref{eq:haahEH} since such terms contribute at third-order in perturbation theory and so can be safely ignored. Nonetheless, this does prevent any local first order contributions to the EH of the form $X_v^1 I_v^2$ for a vertex $v$ in the triangular boundary layer of $B$. While such terms commute with all complete stabilizers in $B$, they do not commute with the superficial NLSS, which is a consequence of the fractal subsystem symmetries present in Haah's code. As we discuss in detail in Sec.~\ref{sec:sspt}, although such \textit{local} terms are disallowed due to the superficial NLSS, there exist {\it quasi-local} combinations of such operators which may yet survive, with consequences for the entanglement structure of the phase. We return to this point in Sections.~\ref{sec:sspt} and~\ref{sec:summary}.

As an aside, we note that forthcoming work by one of us~\cite{albertnew} has found a coupled layer construction of Haah's code, similar in spirit to the coupled layer construction of the X-cube model put forth in Refs.~\cite{han,sagar}. Much like the EH for the X-cube, which clearly distinguishes the fracton phase from decoupled stacks of $d=2$ topological orders, we expect that the EH for Haah's code should likewise be distinct from decoupled stacks of the underlying $d=2$ SSPT layers. We leave an analysis of the effect of the coupling procedure on the entanglement structure to future work.

%Ref.~\cite{albertgauge} conjectured a coupled layer construction for Haah's code, similar to that of Refs.~\cite{han,sagar} which obtains the X-cube model from strongly coupled stacks of the 2d toric code model. For Haah's code, the results of~\cite{albertgauge} suggest that at least one stack of layers must be in the $[111]$ direction and that each $d=2$ layer should be related to the quantum Newman-Moore model, which we recall is an example of an SSPT phase. The EH in Eq.\eqref{eq:111haah} again seem to support this conjecture. However, the results of a forthcoming paper \red{cite}, confirms that Haah's code can be found via coupled stacks of SSPT models, but the models are not Newman-Moore and are not stacked in the $[111]$ direction. Still the $[111]$ direction is important to the procedure, just not in a obvious way. Thus we caution that the EH is a useful hint at the nature of the phase, but the connection can be less than obvious. 

\subsection{Edge-ES Correspondence}
\label{edgeES}

Following the original proposal by Li and Haldane~\cite{lihaldane}, in which the low-energy entanglement spectrum was conjectured to match the low-energy spectrum of physical boundary modes, the edge-ES correspondence and variations thereof were found to hold in many gapped topological phases. For \textit{chiral} topological orders in $d=2$, whose edge excitations are well known to be described by a 1+1D conformal field theory (CFT), it was shown that the reduced density matrix of a bipartition in the bulk is equivalent to the thermal density matrix of the CFT describing the low-energy boundary dynamics~\cite{qi2012}; the edge-ES correspondence was similarly established for a series of fractional quantum Hall states in Refs.~\cite{regnault1,papic,chandran2011,dubail2012,cano2015}. A geometric proof later extended the applicability of the edge-ES correspondence to physical systems with an approximate Lorentz invariance at low energies~\cite{swingle2012}, including topological insulators, which belong to the class of symmetry protected topological (SPT) phases. In SPT states, which are short-range entangled states with non-trivial protected boundary modes, the ES harbours a corresponding symmetry protected degeneracy~\cite{pollmann2010,turner2010,fidkowski2010,alba2012,choo2018}.

While both chiral topological orders and SPT states in $d=2$ host non-trivial gapless edge excitations, with corresponding signatures of these states appearing in the bulk ES, the boundaries of \emph{non-chiral} topologically ordered states, examples of which include the $d=2,3$ toric code, are generically gapped. Nonetheless, it was shown in Refs.~\cite{ho1,ho2,berg2017,stringnetespec} that a version of the edge-ES correspondence holds for such phases as well. Specifically, Ref.~\cite{ho1} studied the Wen-plaquette model, which is unitarly equivalent to the $d=2$ toric code, in the presence of arbitrary local perturbations. While they did not find an exact matching between the low-lying spectrum of the physical edge Hamiltonian and the low-lying ES for generic perturbations, as is true for chiral topological states, a ``weak'' edge-ES correspondence was found to hold. The weaker form of the edge-ES correspondence is encapsulated by the fact that both the EH and the edge Hamiltonian universally belong to the class of $\mb{Z}_2$ invariant $d=1$ Hamiltonians acting on effective spin-$1/2$ degrees of freedom. We now provide evidence that a similar edge-ES correspondence generically holds for the gapped fracton phases considered in this paper as well. 

We first note that for a perturbed stabilizer Hamiltonian~\eqref{eq:ham}, the derivation of the edge Hamiltonian (the Hamiltonian describing a physical boundary of the system), proceeds analogously to that of the entanglement Hamiltonian \textit{i.e.,} through UPT. That is, once a specific physical boundary is specified, we can repeat the perturbative analysis developed in Sec.~\ref{sec:pert} (or equivalently, in Ref.~\cite{ho1}) in order to find $H_{\text{edge}}$ to any given order in perturbation theory. For all models considered in this section, we find that to lowest order, 
\beq
H_{\text{edge}} \simeq H_{\text{ent}},
\eeq
up to shifting and rescaling. Here, $\simeq$ implies that both models belong to the same class of Hamiltonians.

In agreement with the results of Ref.~\cite{ho1}, we find that both $H_{\text{edge}}$ and $H_{\text{ent}}$ for the $d=2$ toric code can be universally mapped onto a $\mb{Z}_2$ invariant, self-dual TFIM acting on effective spin-$1/2$ variables along the one-dimensional entanglement cut. Similarly, for the $d=3$ toric code, both the EH and the edge Hamiltonian can be generically mapped onto an effective $d=2$ TFIM which is invariant under a global $\mb{Z}_2$ symmetry. In fact, it known on general grounds that the boundary theory for a $\mb{Z}_2$ topologically ordered phase in $D$-dimensions is unitarily equivalent to a $\mb{Z}_2$ invariant Ising theory in $D-1$ dimensions~\cite{freed}, where the $\mb{Z}_2$ symmetry is enforced by the bulk topological order. Along with our results in Sec.~\ref{toriccode}, this clearly establishes the edge-ES correspondence for the toric code. Ref.~\cite{freed} additionally shows that a topologically ordered phase with electromagnetic duality in the bulk hosts a boundary theory which is Kramers-Wannier (or self) dual, which we demonstrated was also the case for the EH of the $d=2$ toric code.

For both type-I and type-II fracton orders, typified by the X-cube and the cubic code models respectively, we found that the EH cannot be universally mapped onto some Ising-like model. Instead, it is only for entanglement cuts along the subsystem symmetries (planar for the X-cube, fractal for Haah's code) that the EH can be mapped onto an Ising-like model invariant under a sub-extensive set of $\mb{Z}_2$ subsystem symmetries. The geometric sensitivity of the entanglement Hamiltonian hence serves to clearly distinguish fracton order from topological order, a feature which carries over to the boundary Hamiltonian of fracton phases as well. Curiously, much as the bulk $\mb{Z}_2$ topological order enforces global $\mb{Z}_2$ invariance in the edge and entanglement Hamiltonians for the toric code, the bulk $\mb{Z}_2$ fracton order enforces a subsystem $\mb{Z}_2$ invariance on the edge and entanglement Hamiltonians for both the X-cube and the cubic code. For the EH, this invariance is enforced through the NLSS, while for the edge Hamiltonian we expect that it is a consequence of the Wilson/'t Hooft operators in the bulk. Based on our results, it seems reasonable that results established in Ref.~\cite{freed} for topological order can be extended to gapped phases with fracton order as well. In other words, we expect that the boundary Hamiltonian (equivalently, EH) for a system with fracton order can be mapped onto a subsystem symmetric Ising model, as long as the physical boundary (entanglement cut) lies along the planar (for type-I) or fractal (for type-II) subsystem symmetries present in the bulk. While boundary theories of fracton phases have yet to receive much attention, we note that they have been studied in some detail for the X-cube model in Ref.~\cite{bulmashboundary}, whose results match ours where there is overlap.

%%%%%%%%%%%%%%%%%%%%%%%%%%%%%%%%%%%%%%%%

\section{Quasi-Local Entanglement in SSPT Phases}
\label{sec:sspt}

\begin{figure}[t]
    \centering
    \includegraphics[scale=.5]{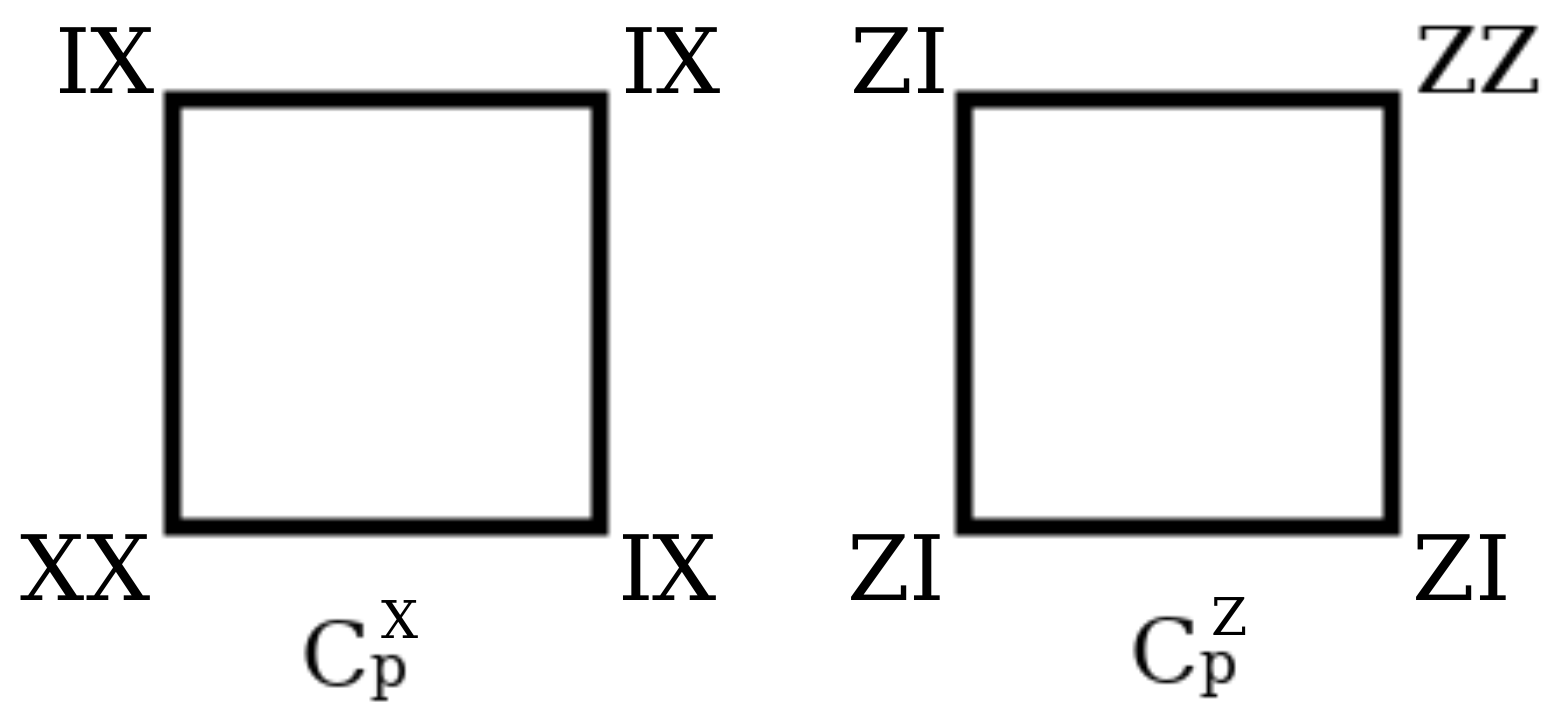}
    \caption{Definition of the local stabilizers for the CSS variant of the cluster model.}
    \label{fig:ssptdef}
\end{figure}

In this section, we look at the entanglement Hamiltonian of stabilizer codes describing subsystem symmetry protected topological phases~\cite{yizhi1,yizhi2,devakulfractal,strongSSPT}. In particular, we consider the $d=2$ cluster model, using a CSS variant of the coarse-grained version introduced in Ref.~\cite{spurious}\footnote{Our version differs from that discussed in Ref.~\cite{spurious} by a local unitary which applies the Hadamard gate on the second qubit for every vertex.}. This model is defined on the square lattice with two qubits assigned to each vertex and the Hamiltonian
\beq
H_{cl}= - \sum_p \left(C_p^X + C_p^Z\right),
\eeq
where $C_p^X, C_p^Z$ are given by translates of the terms defined in Fig.~\ref{fig:ssptdef}.

In Ref.~\cite{spurious} it was shown that this model, which is an example of a strong SSPT phase (a notion defined in Ref.~\cite{strongSSPT}), harbours contributions to the topological entanglement entropy for square entanglement cuts along the coordinate directions. Ref.~\cite{spurious} dubbed such contributions ``spurious'' since they survive the usual subtraction scheme defining the TEE~\cite{preskill,levin2006}, which is expected to vanish identically in gapped phases without long-range entanglement \textit{i.e.,} without topological order. Unlike contributions to the TEE for LRE states, such as the toric code, which are independent of the entanglement cut, the spurious contributions for the cluster state instead only arise for square cuts. This suggests that these spurious contributions arise as a consequence of superficial NLSS associated with the subsystem symmetry protecting the state. These subsystem symmetries are given by the product of all stabilizers of a single type along any rigid line in either coordinate direction. The corresponding operators are
\begin{align}
g_{1y}^X = \prod_x (XI)_{(x,y)}\nonumber\\
g_{2x}^X = \prod_y (XI)_{(x,y)} \nonumber\\
g_{1y}^Z = \prod_x (IZ)_{(x,y)}\nonumber \\
g_{2x}^Z = \prod_y (IZ)_{(x,y)},
\end{align}
where $(x,y) \in \mb Z_L^{2}$ indexes a vertex of the graph. We now focus on the EH for square cuts to understand the significance of the subsystem symmetries, which we argue endow the strong-SSPT phase with ``quasi-local" entanglement.

Due to the lack of any rotational symmetry in this model, each straight edge of the square entanglement cut has slightly varying contributions to the EH. For the top and left boundary, we find $(3,2)$ cut $X$ stabilizers and $(2,3)$ cut $Z$ stabilizers, whereas for the bottom and right boundary we find $(3,2)$ cut $Z$ stabilizers and $(2,3)$ cut $X$ stabilizers. At the corners of the cut, there is one $X$ and one $Z$ cut stabilizer, which are either $(2,3)$ or $(1,4)$ depending on the corner. For each edge along the cut, we find that the EH maps onto the $d=1$ TFIM, which is also what we would find for a square entanglement cut for the $d=2$ toric code (see Sec.~\ref{toriccode}). However, the transverse field terms of the effective TFIM do not get projected onto a single charge sector, unlike the analogous terms for the toric code EH. 

Crucially though, the nearest-neighbour Ising terms of the effective EH satisfy a ``zero-charge'' condition. This is attributed to the superficial NLSS given by the product of cut stabilizers in the set defining one of the subsystem symmetries along that boundary. As a consequence of the subsystem symmetry, such NLSS are concomitant with any rigid entanglement boundary, with the subsystem invariance hence fundamentally responsible for the survival of spurious contributions to the TEE. Surprisingly, it appears that the subsystem symmetries are behaving much as topological constraints do, resulting in contributions to the entanglement entropy which are typically associated with LRE systems. This is in consonance with recent work suggesting that, as a consequence of subsystem symmetries being intermediary between global and local (gauge) symmetries, strong SSPT phases may have a robust ground state degeneracy on topologically non-trivial manifolds~\cite{hughessspt}. Hence, strong SSPT phases are distinct from their SRE SPT counterparts and also from LRE topologically ordered phases, instead harbouring a pattern of \textit{``quasi-local'' entanglement (QLE)} in their ground states.

More precisely, we can view the subsystem symmetries as enforcing a psuedo-constraint in the following way: suppose we impose a {\it restriction} on the kinds of Pauli operators which are allowed to act on our system. Viewed as a limitation on the types of perturbations which are allowed, these must form a subspace of the Pauli space such that products of any of the restricted operators also belong to the set of restricted operators. In this case, let us now consider all operators which commute with the subsystem symmetries of the $d=2$ cluster states (modulo any of the stabilizers). Such a subspace $\mc R \subseteq P$ is generated by the set $\{XI, IZ\}$ for every vertex of the square lattice. Due to the restriction, not all parts of the stabilizers matter, \textit{i.e.,} we can project out the parts of the operators which commute with everything in $\mc R$, which happens to be $\mc R$ itself. As a result, the stabilizer set becomes two commuting copies of the Ising plaquette model and, significantly, the subsystem symmetries are projected onto the identity after the restriction. This implies that each subsystem symmetry imposes a psuedo-$\mb{Z}_2$ conservation law once the restriction is enforced. This restriction is also reflected in the EH.

N\"iavely, one might think we have left out first-order contributions to the EH coming from $IX$ and $ZI$ perturbations. But while these terms locally commute with all complete stabilizers, they do not commute with the subsystem symmetry and thus do not commute with the superfical NLSS. However, such terms can contribute if, upon inclusion of their coefficients, they collectively commute with the superficial NLSS {\it i.e.,} act \textit{quasi-locally}. In other words, we can construct an operator $\mc{O} = \sum_i \chi_i \mc{O}_i$ which commutes with the superficial NLSS despite being composed of local terms $\mc{O}_i$ which do not individually commute with the superficial NLSS. In order for this to occur, the coefficients $\chi_i$ must conspire in a manner which preserves the invariance of $\mc{O}$ under the subsystem symmetries, implying a measure of non-locality in the system. In the language of Ref.~\cite{spurious}, the existence of such an operator $\mc{O}$ implies the absence of a \textit{linearly-symmetric} local unitary transformation which can transform the cluster state into a trivial product state, thereby demonstrating the strong SSPT nature of the cluster state. In fact, the quasi-local nature of the strong SSPT phase is implicit in the classification developed in Ref.~\cite{strongSSPT}: the unitary circuit required to transform a strong SSPT into a trivial product state, while of finite depth, must have support over a \textit{sub-extensive} number of sites \textit{i.e.,} is ``infinite-length'' in the thermodynamic limit. This is in contrast with both SPT states, for which the corresponding unitary would have support over a finite number of sites, and with topologically ordered states, for which it would need support over an extensive number of sites.

Thus, we see that strong SSPT states are quasi-locally entangled, since they possess features characteristic of long-range entangled states, but only for entanglement cuts and perturbations respecting the subsystem symmetry: for a square entanglement cut, as long as the perturbations respect the subsystem symmetry, the EH is equivalent to the $d=1$ TFIM. If we allow perturbations which break these ``gauge-like'' symmetries~\cite{ortiz}\footnote{A gauge-like symmetry is one which acts independently on a $d$-dimensional sub-region of a $D$-dimensional system, with $0<d<D$.}, we generically find additional non-local terms in the EH, destroying its mapping onto the TFIM. In contrast with the EH for the $d=2$ toric code, which is equivalent to the $d=1$ TFIM for generic perturbations, the EH for the $d=2$ cluster state only maintains this equivalence for a certain class of perturbations; equivalently, the mapping of the EH onto the $d=1$ TFIM is protected by the subsystem symmetry. Hence, we see that the EH for the cluster state resembles that for a topologically ordered state but only as long as the subsystem symmetry is preserved. Since the entanglement entropy can be extracted from the EH, this further implies that the so-called spurious contributions to the TEE in the strong SSPT phase are in-fact relaying the quasi-local entanglement present in the state. It is in this sense that we assert that these seemingly ersatz contributions to the TEE---although not topological in nature---possess a measure of robustness beyond that found in conventional SPT phases and are indicative of the gauge-like nature of subsystem symmetries.

%%%%%%%%%%%%%%%%%%%%%%%%%%%%%%%%%%%%%%%%

\section{Discussion: Role of NLSS}
\label{sec:summary}

The primary result of Section~\ref{ES}, which is the derivation of the EH Eqs.~\ref{eq:EntHA} and \ref{eq:EntHB} for stabilizer codes in the presence of arbitrary (weak) local perturbations, leads to several interesting entanglement Hamiltonians, as we have established in the preceding sections. As a corollary, we have also provided evidence that a weak edge-ES correspondence, known to hold for non-chiral topologically ordered states~\cite{ho1,ho2,berg2017,stringnetespec}, also extends to systems with fracton order. Another key result, which we now expound upon, is the existence of non-trivial features in the entanglement structure which are protected by the NLSS. 

Thus far, all of our results have been established for the ground states of the perturbed systems. To extend this, we conjecture that the EH for low-energy, perturbed eigenstates takes the same form as that for the ground state. More precisely, we posit that the EH for a low-energy eigenstate of the perturbed system is also a sum of projected cut stabilizers, but with coefficients which may differ from those for the ground state. The validity of this conjecture comes down to evaluating the validity of the UPT unitary acting on the low-energy states. That is, if we calculate Eq.~\eqref{eq:split} with $U$ acting on a low-energy stabilizer eigenstate $\ket{\mbf{k}}$, the analysis proceeds identically (up to the exact coefficients) so long as the excitations do not lie near the boundary defined by the cut. An important consequence of this is that for the emergent Gauss law $\prod_{i \in F_A} O_i =\left(g_\partial\right)_A$ from Eq.~\eqref{eq:gauss}, Eq.~\eqref{eq:constraint} becomes
\beq\label{eq:blkbdy}
\prod_{q \in F_\partial }\tilde X_q = (-1)^c \, P_A^{(\bk_A)},
\eeq
where $F_\partial$ is the set of stabilizer indices such that $g_\partial =\prod_{i \in F_\partial} O_i$ and $c$ represents the number of excitations contained in $F_A$ modulo 2 for the state $\ket{\mbf{k}}$. This implies the surface charge of the EH is confined to the $c$-charge sector. If this holds, the charge of the surface EH can measure the charges in the bulk, which constitutes a ``charge'' bulk-edge correspondence. Here, the charge is defined by a global conservation law enforced by a topological constraint, which also enforces the existence of an NLSS (see Refs.~\cite{albertgauge,albert}).

However, all excited states of the models considered here can have additional {\it non-topological} degeneracy {\it i.e.,} a degeneracy which is not related to logical operators. As a result, it is not clear that we can ignore the choice of basis within the excited degenerate eigenspaces\footnote{In UPT, this translates into the application of a constant (in $\lambda$) unitary which is block diagonal in the degenerate subspaces, but has no matrix elements between different eigenstates of the logical operators. Because of this logical operator selection rule, the ground space remains unaffected by the change of basis operator and we are hence justified in ignoring any such operators in our original analysis.}. An intuitive argument for why one should be allowed to safely ignore non-topological degeneracies is as follows: in the original definition of the stabilizer Hamiltonian Eq.~\eqref{eq:ham}, we allowed for arbitrary coefficients $\{J_s\}$. Aside from modifying the final coefficients appearing in the EH, allowing arbitrary coefficients $\{J_s\}$ does not affect the results of Sections~\ref{sec:unpert} and \ref{sec:pert}, but can remove any non-topological degeneracies from the excited states. Stated differently, simply by choosing a subset of the coefficients $\{J_s\}$ to be negative, we can make any valid stabilizers' eigenstate the unperturbed ground state uniquely, up to the topological degeneracy.

%Even with the simple adaptation of selectively choosing a subset of $\{J_s\}$ to be negative, we can make any valid stabilizers eigenstate the unperturbed ground state uniquely up to topological degeneracy. 

%Assuming our conjecture holds for some range of low-energy, perturbed eigenstates, it is then straightforward to find the emergent Gauss' Law constraints given by the NLSS in the ES as follows. For an eigenstate labelled by $\bk$, such that $c$ charges exist in the bulk of region $A$ (\AP{\textit{i.e.,} away from the entanglement cut}), the projection of the cut stabilizers in the EH is given by $P_A^{(\bk_A)}$. In this case, Eq.~\eqref{eq:constraintHere, the charge is defined by a global conservation law enforced by a topological constraint, which also enforces the existence of an NLSS (see Refs.~\cite{albertgauge,albert}).} is replaced by 
%\beq
%\prod_{q \in F\subseteq \partial A}\tilde X_q = (-1)^c \, P_A^{(\bk_A)},
%\eeq
%implying that the surface charge sector must correspond to the $\mb{Z}_2$ number of charges in the bulk. \AP{Here, the charge is defined by a global conservation law enforced by a topological constraint, which also enforces the existence of an NLSS (see Refs.~\cite{albertgauge,albert}).}
%By ``charge", we are here referring to a topological constraint which both implies a global conservation law that defines the charge (see Ref.~\cite{albertgauge}), as well as the existence of an NLSS (see Ref.~\cite{albert}).

Assuming this conjecture holds, the topological constraint is fundamentally responsible for enforcing the correspondence between the surface charge sector and the number of bulk $\mb{Z}_2$ charges, evidence that such a charge bulk-edge correspondence is endowed with a measure of topological invariance. For topologically ordered (LRE) gapped phases, this bulk-edge correspondence has a strict topological invariance, in that it is robust against any local perturbations and appears universally for any entanglement cut. Moreover, this conjecture also implies that flux conservation presents clear signatures in the ES for $d=3$ topological orders, such as the $d=3$ toric code model. If the entanglement boundary has non-trivial topology, then one can measure a surface flux analogous to Eq.~\eqref{eq:blkbdy}, corresponding to the number of fluxes enclosed or encircled by the cut. This is a consequence of the NLSS being associated with flux and charge topological constraints in the bulk. On the other hand, for strong SSPT phases such as the $d=2$ cluster state, the bulk-edge correspondence is robust only for entanglement cuts and perturbations respecting the subsystem symmetry. For gapped systems with fracton order, there exists a mesh of both kinds of invariance (topological and subsystem), since although the total number of charge types is sensitive to the geometry, a subset of these are topological in nature \textit{i.e.,} exist universally for arbitrary local perturbations and for generic cuts.

Where SSPT phases and fracton systems differ from topologically ordered phases is in the existence of NLSS which %are not associated with any topological constraints but which instead stem directly from the subsystem symmetries;
are not invariant under all equivalent bulk transformations of $A$, but are only invariant under transformations within the intersection of $A$ with the subsystem symmetries. This is evidenced by the existence of superficial NLSS when the cut aligns with the rigid subsystems. For SSPTs these subsystem symmetries are not strictly topological, but rather quasi-local in nature (see previous section), while for fractonic systems their nature may be either topological (as is the case for the X-cube superficial NLSS) or quasi-local (as is the case for the fractal superficial NLSS in Haah's cubic code). %In either case, these additional NLSS are superficial since one expects that the surface along which the superficial NLSS forms corresponds to the subsystem symmetry.
Though the quasi-local superficial NLSS can not be associated with a bulk-edge topological charge correspondence, such NLSS can nevertheless be attached to a ``pseudo-constraint'' provided by the subsystem symmetry and a \textit{restriction} on the Pauli space. Here, a restriction refers to a limitation on the kind of Pauli operators which are allowed to act on a subspace, whereby the type of excitations are limited as well. Restrictions can be formalised in the language of \textit{linear gauge structures}, developed in Ref.~\cite{albertgauge}. In this way, a subsystem symmetry which is contained in $G$ implies a psuedo-charge if there exists a restriction such that the subsystem symmetry is elevated to a constraint. Thus, under the restriction, the subsystem symmetry implies a conservation law, as demonstrated explicitly for the SSPT example in Sec.~\ref{sec:sspt}. It was in this sense that we asserted that the seemingly spurious contributions to the entanglement, despite not being topological in origin, are a consequence of the subsystem symmetry and imply constraints analogous to those typically encountered in LRE states. The presence of such NLSS endows the strong SSPT with a measure of quasi-local entanglement beyond that encountered in conventional SPT states but lacking the full topological protection associated with topologically ordered states. Thus, the lexicon of gapped quantum phases must be expanded to include QLE states as intermediaries between those with either short- or long-range entanglement.

%As for SSPT and fracton system (at least our Type-II example) there are NLSS which are not linked to an constraint, but rather directly to the subsystem symmetry. In all cases here, these are superficial NLSS and should be generally true as one expects the surface along which the superficial NLSS forms corresponds to the subsystem. Even though the NLSS is not attached to a topological charge, it can be attached to a pseudo-charge provided by the subsystem symmetry and a {\it restriction} on the Pauli space. A restriction refers to putting a limitation of the type of Pauli operators which act on a subspace, whereby the type of excitations are then limited as well. Restrictions can be formalized in the language of {\it linear gauge structures} (see Ref.~\cite{albertgauge}) which will be made explicit in future work. So a subsystem symmetry which is contained in $G$ implies a psuedo-charge if there exists a restriction such that the subsystem symmetry becomes a constraint. Thus under the restriction, the subsystem symmetry implies a conservation law\SJ{, as demonstrated explicitly for the SSPT example in Sec.~\ref{sec:sspt}.} It is in this sense that we assert that these so-called ``spurious'' contributions to entanglement coming directly from superficial NLSS attributed to the subsystem symmetry are important, even if not formally topological.

 %Relation to Ref.~\cite{brandao2018}?

%%%%%%%%%%%%%%%%%%%%%%%%%%%%%%%%%%%%%%%%

\section{Conclusions and Outlook}
\label{cncls}

In this paper, we have discussed the entanglement spectrum for the ground state of a general stabilizer Hamiltonian, first by considering the unperturbed Hamiltonian and then by analysing the effects of weak, local perturbations via UPT. This led us to find a version of the edge-ES correspondence between the entanglement Hamiltonian in the bulk and the edge Hamiltonian along the boundary of the system. More importantly, we found universal entanglement features which are preserved under such generic perturbations as enforced by NLSS symmetries of the stabilizer Hamiltonian. These features take the form of emergent surface charges/fluxes, which, via the NLSS and the emergent Gauss' law it implies, are related to charges/fluxes in the bulk. We interpret this as a type of ``charge'' bulk-boundary correspondence. We applied our general results to several examples, including conventional topological order as encapsulated by the $d=2,3$ toric codes, type-I and type-II fracton order as encapsulated by the X-cube model and Haah's cubic code, and strong SSPT phases as encapsulated by the $d=2$ cluster model.

Whereas the emergent surface charges/fluxes were found to possess a robust topological invariance for conventional topological order, both fracton and strong-SSPT phases demonstrate a non-trivial dependence on geometry, corresponding to their subsystem symmetries. Our results hence establish the ES as a clear diagnostic for not only fracton order, but also for strong SSPT phases, with the latter characterised by a qualitatively distinct pattern of quasi-local entanglement. This suggests that SSPTs, despite not having the robust topological invariance associated with LRE gapped states, exhibit entanglement features beyond those expected for familiar SPT phases, as reflected in the invariance of their ES under subsystem preserving perturbations.

We also conjectured that the surface charges/fluxes could be used to measure the bulk charges/fluxes for perturbed excited states so long as certain assumptions about UPT hold for these states. In principle, one should be able to place bounds on the range of low-energy excited states for which we expect the conjecture to hold; we leave a such a detailed analysis to future work. Further, in this paper we have restricted our attention to understanding the ES of fracton and SSPT phases described by stabilizer code Hamiltonians. It would be interesting to go beyond the stabilizer formalism and study the ES for fracton orders not described by commuting-projector Hamiltonians. While the edge-ES correspondence has been established more generally for $d=2$ non-chiral, non-Abelian topological orders with gapped boundaries~\cite{stringnetespec}, it remains to be seen whether similar results hold for $d=3$ non-Abelian fracton models with gapped boundaries~\cite{sagar,cagenet,twisted}. As another extension, we note that although commuting-projector Hamiltonians cannot describe chiral phases~\cite{kitaev}, it has been suggested that gapped chiral fracton phases can instead be captured by the language of higher-rank tensor gauge theory~\cite{chiral,prem2}. In the future, it would be interesting to study whether a stronger edge-ES correspondence, similar to that for chiral topological orders, exists for chiral fracton phases as well. More generally, understanding the structure of entanglement in gapless fracton phases is an important future direction, albeit one which lies beyond the techniques developed in this paper.

\begin{acknowledgments}
The authors are especially grateful to Rahul Nandkishore for inspiring discussions and prior collaborations on fractons. We also acknowledge stimulating conversations with Danny Bulmash, Meng Cheng, Michael Hermele, Wen-Wei Ho, Tom Iadecola, Michael Pretko, Hao Song, Ruben Verresen, Dominic Williamson, and Han Yan. AS is supported by the Air Force Office of Scientific Research under award number FA9550-17-1-0183. SJ acknowledges support by the U.S. Department of Energy, Office of Science, Basic Energy Sciences (BES) under Award number DE-SC0014415. AP is funded by a PCTS Fellowship at Princeton University.

\end{acknowledgments}

%%%%%%%%%%%%%%%%%%%%%%%%%%%%%%%%%%%%%%%%
%%%%%%%%%%%%%%%%%%%%%%%%%%%%%%%%%%%%%%%%

\appendix

\section{Unitary Perturbation Theory}
\label{UPT}

We discuss our method for perturbatively constructing a unitary operator $U(\lambda)$ which relates eigenvectors of an unperturbed Hamiltonian to the dressed eigenvectors of a perturbed Hamiltonian. Our derivation, which is a variation on the usual Schrieffer-Wolff perturbation theory~\cite{swpert,bravyisw}, is designed to preserve the unitarity of $U(\lambda)$ to all finite orders in perturbation theory and is closely related to the Wegner-Wilson flow discussed in the context of many-body localization~\cite{imbrie,pekker2017}.

Consider the case where one has an unperturbed Hermitian operator $H_0$ acting on members of a finite-dimensional Hilbert space $\Hi$ with eigenvalues $\{E_n\}_{n \in I}$ and a corresponding complete orthonormal set of eigenvectors $\{\ket{n}\}_{n \in I}$. Note that this set is not unique in the case of a degenerate spectrum. Now, suppose we add a Hermitian perturbation $V$, weighted by a control parameter $\lambda \in \R$, such that our new Hermitian operator becomes
\beq
H(\lambda) = H_0 + \lambda V.
\eeq
We now wish to approximate the new eigenvalues $\{E_n^\prime\}_{n \in I^\prime}$ and eigenbasis $\{\ket{n^\prime}\}_{n \in I^\prime}$ of the perturbed Hamiltonian at any finite order in $\lambda$. As discussed above, we can achieve this by finding a unitary $U(\lambda)$ which maps $\{\ket{n}\}_{n \in I}$ to $\{\ket{n^\prime}\}_{n \in I^\prime}$. However, we must define the mapping between the two bases, in order for which we make the following assumptions:
\begin{enumerate}
\item $\ket{n^\prime(\lambda)}$ is an eigenvector of $H(\lambda)$,
\item $\ket{n^\prime}$ is analytic within some non-zero radius of convergence about $\lambda =0$,
 \item $\lim_{\lambda\to 0} \ket{n^\prime (\lambda)} = \ket{n^\prime(0)} =\ket{n}$, and 
\item for any other $\ket{m} \in \{\ket{n}\}_{n \in I}$, its associated function $\ket{m^\prime(\lambda)}$ satisfies $\braket{n^\prime(\lambda)|m^\prime (\lambda)} = \delta_{nm}$, for all $\lambda$ within the radius of convergence. 
\end{enumerate}

Taken together, these conditions imply that there exists an analytic, operator-valued function of $\lambda$, $U(\lambda)$, which transforms the unperturbed eigenbasis into the new eigenbasis. Further, this implies the existence of an analytic skew-Hermitian operator-valued function of $\lambda$, $F(\lambda)$, such that $U(\lambda) = \exp(F(\lambda))$. We now define the operator-valued function
\beq
\label{eq:Ddef}
D(\lambda) = U(\lambda) H(\lambda)U(\lambda)^\dagger,
\eeq
which is also analytic. We can expand both sides of this equation using the expansions
\begin{subequations}
\begin{align}
D= \sum_n D^{(n)} \lambda^n,\\
F= \sum_n F^{(n)} \lambda^n,
\end{align}
\end{subequations}
and match the two sides, order by order. Rearranging Eq. \eqref{eq:Ddef}, one has that
\begin{align}
H_0 + \lambda V = U^\dagger D U = \exp(i\ad_F) D, \label{eq:interm}
\end{align}
where for any operators $A$ and $O$, $\ad_A O = [A,O]$. Now we can expand both sides in $\lambda$ and equate coefficients at each order in $\lambda$. The LHS is already expanded, whereas the RHS is a bit more complicated:
\begin{widetext}
\begin{align}
\exp(i\ad_F) D =& \sum_n \frac{1}{n!} \ad_F^n D \nonumber \\
=& \sum_{n,m}\frac{1}{n!} \sum_{a_1} \sum_{a_2} \dots \sum_{a_n} \lambda^{\|\vec{a}\| + m} [F^{(a_1)}, [ F^{(a_2)},[ \dots [ F^{(a_n)}, D^{(m)}] \dots ]]] \nonumber \\
=& \sum_n \lambda^n \sum_m \mathcal F ^{(n-m)} D^{(m)},
\end{align}
\end{widetext}
where $\|\vec{a}\| = \sum_i a_i$ and where the super operator $\mathcal F^{(n)}$ is defined as acting as the identity for $n=0$ and, for $n>0$, having the action on any operator $O$:
\begin{align}
\mathcal F^{(n)} O = \sum_\alpha \sum_{\vec{a} \in \mb N^{\otimes \alpha}}^{\|\vec{a}\|=n} \frac{1}{\alpha!}[F^{(a_1)}, [F^{(a_2)},[ \dots [ F^{(a_\alpha)}, O] \dots ]].
\end{align}
We then find the following relations, order by order:
\begin{subequations}\label{eq:interm2}
\begin{align}
\lambda^0:&\, H_0 = D^{(0)},\\
\lambda^1:& \, V = \mc F^{(1)} D^{(0)} + D^{(1)}= D^{(1)} - [H_0, \mc L],\\
\vdots \nonumber \\
\lambda^n:& \, 0 = \sum_{m} \mc F^{(n-m)} D^{(m)},
\end{align}
\end{subequations}
where $\mc{L}:=F^{(1)}$. To solve for each operator, we define the super-operator $\int_{H_0}$, with an action on an operator $O$ of   
\beq
\int_{H_0} O = \sum_{n,m \in I} \frac{\braket{n|O|m}[E_n \neq E_m]}{E_n -E_m} \outerp{n}{m}.
\eeq
The super-operator $\int_{H_0}$ is a partial inverse to $\ad_{H_0}$ up to an arbitrary operator which commutes with $H_0$. Furthermore, the kernel of this function is all operators commuting with $H_0$ which, by definition, includes all $D^{(n)}$. Hence, at order $n$, we apply the $\int_{H_0}$ super-operator to both sides of Eq.~\eqref{eq:interm2}, where we note that the $m=0$ term is $D^{(n)}$, and so it drops out from the sum. Likewise, the $m=n$ term in the sum is $\mc F^{(n)}H_0$, which includes the term $- \ad_{H_0} F^{(n)}$, which maps to $-F^{(n)}$ under $\int_{H_0}$. Since all other terms only involve lower-order operators, one can recursively solve for all $F^{(n)}$ and $D^{(n)}$. At lowest order, we find that
\beq
\mc{L} = -\int_{H_0}V= -\sum_{n,m \in I} \frac{\braket{n|V|m}[E_n \neq E_m]}{E_n -E_m} \outerp{n}{m},
\eeq
as quoted in Eq.~\eqref{eq:Lstab} in the main text for the stabilizer Hamiltonian. Note that all $F^{(n)}$ are not unique as we are always allowed to add an operator which commutes with $H_0$ at each step of the recursion. Above, we are implicitly choosing this operator to be zero. 

\section{Canonical Basis for the Pauli Group} \label{ap:canon}

In this Appendix, we describe the notion of a canonical basis for the Pauli group over $N$ qubits. As the Pauli group has been ``Abelianized,'' we can now treat it as a vector space over $\mb Z_2$, where addition is given by the product of Pauli operators and scalar multiplication is given by the power of a Pauli operator. However, it appears that we have lost the commutation rules for Pauli operators. We can recover this by introducing the symplectic form $\lambda: P \times P \to \mb Z_2$, whereby for $p_1, p_2 \in P$, $\lambda(p_1,p_2)=1$ if the two operators anti-commute, and is zero otherwise. 

We start by considering the trivial example of a canonical basis as given by $\{X_i, Z_i\}_{i<N}$. Clearly, this is a basis for $P$, but to use it along with $\lambda$ to expand any Pauli operator (in analogous fashion to an othonormal basis in an inner product space), one observes that this basis has the properties $\lambda(X_i, X_j)= \lambda(Z_i,Z_j) =0$ and $\lambda(X_i, Z_i)= \delta_{ij}$. Thus, one can expand any Pauli $ p \in P$ as 
\beq \label{eq:pauliex}
p =\sum_i \lambda(p,X_i) Z_i + \sum_i \lambda(p,Z_i)X_i.
\eeq
If we wish to have a similar expansion in some other basis, we demand a similar set of requirements. We define a canonical basis as a spanning set  $\{p_i\}_{i<N} \cup \{\tilde p_i\}_{i<N}$ such that $\lambda(p_i, p_j) = \lambda(\tilde p_i, \tilde p_j)=0$ and $\lambda(p_i, \tilde p_j)= \delta_{ij}$. These conditions imply that we have an expansion similar to Eq.~\eqref{eq:pauliex} for every Pauli operator within this basis. As the stabilizer group $G$ is mutually commuting, any basis for it suffices as (nearly) one-half of a canonical basis, but the other half, a canonical dual, is only unique up to members of $G$. This also suffices to show the dimension of any mutually commuting subspace of $P$ is maximally $N$. 

\section{Recoverable Information for the $[111]$ cut in Haah's code}
\label{ap:RI}

Here, we derive the recoverable information for an $R\times R \times R$,  $45 \degree, 45 \degree, 60 \degree$ $[111]$ parallelepiped entanglement cut, as illustrated in Fig.~\ref{fig:111cut}. We use the definition
\begin{align}
\mu = d_\partial - s_A- s_B = d_\partial -2 s,
\end{align}
where the last equality is a consequence of using pure states. To compute $s$, we use $s=s_A =|A|- d_G$ and the fact that for bounded simple cuts, all of $G_A$ is generated by stabilizers in $A$~\cite{han2}. Note that we have also removed the minimization over all possible generating sets (see Eq.~\eqref{eq:recover}) as there are no trivial constraints for Haah's code. So as to visually facilitate the derivation, we note that we can visualise this cut as a stack of triangular lattice layers, each of which is off-set with respect to the others such that if we start at a vertex and proceed upwards, we pass through a triangle of the layer immediately above and an oppositely oriented triangle in the layer above that one. This combination of a vertex and two triangles is also one of the more illuminating ways of visualising the stabilizers of Haah's code. If we consider the $X$ stabilizer as ``upwards-pointing'' since it follows the order: vertex $\leftarrow$ triangle $\leftarrow$ triangle, then the $Z$ type stabilizer is ``downwards-pointing'' since it follows the opposite order: triangle $\leftarrow$ triangle $\leftarrow$ vertex. Note that for each triangle forming a stabilizer, the top (bottom) triangular layer is only supported in one of the two qubits at a vertex, the middle triangular layer is supported on the remaining qubits, and the vertex layer is supported on both. We refer to the position of the stabilizer as that of the single vertex.

Clearly, the number of qubits in $A$ is $|A|=2R^3$. As for complete stabilizers, the top two layers contain no complete X stabilizers while the bottom two layers contain no Z complete stabilizers. In every other layer, all stabilizers are complete except for along the two edges near an acute angle of the parallelepiped. This implies $d_{G_A} = 2(R-2)(R-1)^2$ such that $s= 8R^2 -10R+4$. 
As for cut stabilizers, there are clearly $8R^2-10R+4$ coming from all vertices in $A$ which were not counted while calculating $d_{G_A}$. Every triangle in the bottom layer, as well as all those along the edge, represents a cut $X$ stabilizer located in the two layers just below the surface and likewise for the top layer and cut $Z$ stabilizers. This contributes $4R(R+2)$ cut stabilizers. For the remaining layers, every upwards-pointing triangle along an obtuse-angle boundary of the parallelepiped is an additional cut stabilizer. This contributes $2(R-1)(2R-1)$. Applying the definition, we find that
\beq
\mu_{\text{Haah}}^{[111]}= 12R-2.
\eeq
Curiously, this is the same value as that for a cubic cut.  

%---------------------------------------
%BIBLIOGRAPHY
%---------------------------------------

\newpage 

\bibliography{library}

%---------------------------------------
%---------------------------------------
\end{document}